\documentclass[useAMS,usenatbib]{mn2e}

\usepackage{graphicx}

\usepackage{epsfig}

\usepackage{rotating}
\usepackage{aas_macros}
\usepackage{fixltx2e}

\newcommand{\kms}{\mbox{km s$^{-1}$}}
\newcommand{\lsun}{\mbox{${\rm L}_\odot$}}
\newcommand{\msun}{\mbox{${\rm M}_\odot$}}
\newsavebox{\rotbox}

\title[NGC\,765 - A disturbed H\,I giant]{NGC\,765 - A disturbed H\,I giant}
\author[A. M. Portas et al. ]{A. M. Portas$^{1,2,}$\thanks{E-mail: A.M.P.Portas@herts.ac.uk}, E. Brinks$^{1}$,  M. E. Filho$^{2}$, A. Usero$^{3}$, E. M. Dyke$^{1}$ and P.--E. Belles$^{1}$\\
$^{1}$Centre for Astrophysics, Science \& Technology Research Institute, University of Hertfordshire, Hatfield, Herts AL10 9AB \\
$^{2}$Centro de Astrof\'isica da Universidade do Porto, Rua das Estrelas s/n, Porto, Portugal \\
$^{3}$Observatorio Astron\'omico Nacional, C/ Alfonso XII 3, 28014 Madrid, Spain}

\begin{document}

\date{}

\pagerange{\pageref{firstpage}--\pageref{lastpage}} \pubyear{2002}

\maketitle

\label{firstpage}

\begin{abstract}
We present  \textsc{H\,i} spectral line and radio--continuum VLA  data of the galaxy NGC\,765, complemented by optical  and  \textit{Chandra} X-ray maps. NGC\,765  has the largest \textsc{H\,i}-to-optical ratio known to date of any spiral galaxy and one of the largest known \textsc{H\,i}  discs in absolute size with a diameter of ~240 kpc measured at a surface density of $2 \times 10^{19}$ atoms cm $^{-2}$. We derive a  total \textsc{H\,i} mass of M$_{\rm {H\,I}}$ = 4.7$\times 10^{10}$ \msun, a dynamical mass of M$_{\rm dyn} \sim 5.1 \times 10^{11}$ \msun\ and an \textsc{H\,i} mass to luminosity ratio of M$_{\rm{H\,I}}$/L$_{\rm B}=1.6$, making it the nearest and largest `crouching giant'. Optical images reveal evidence of a central bar with tightly wound low-surface brightness spiral arms extending from it. Radio--continuum (L$_{\rm 1.4\,GHz} = 1.3 \times 10^{21}$ W Hz$^{-1}$) and X--ray (L$_{\rm X}$ $\approx$ 1.7 $\times$ 10$^{40}$ erg s$^{-1}$) emission is found to coincide with the optical  core of the galaxy, compatible with nuclear activity powered by a low-luminosity AGN. We may be dealing with a galaxy that has retained in its current morphology traces of its formation history. In fact, it may still be undergoing some accretion, as evidenced by the presence of \textsc{H\,i} clumps the size $<$ 10 kpc and mass (10$^8$--10$^9$ \msun) of small (dIrr) galaxies in the outskirts of its \textsc{H\,i}  disc and by the presence of two similarly sized companions. 

\end{abstract}


\section{Introduction}
\label{Introduction}

The Universe is populated with galaxies spanning a  range of sizes and masses: on the one hand there are gas--rich systems, ranging from  small, amorphous dwarfs to grand--design spirals. On the other, we find from gas--poor objects, from dwarf spheroidals and ellipticals to massive cD galaxies. Their  current morphology and composition contains fossil evidence regarding their formation and evolution \citep{fre02}. Accretion of small clumps of both baryonic and dark matter, falling towards more massive  dark matter haloes, is now a well accepted scenario for the assembly of late--type galaxies \citep{nav95, spr05}. Discs form by cooling  and the dissipative collapse of baryonic matter \citep[e.g.,][]{fal80}. During this process, infalling gas should preserve a large fraction of its angular momentum whilst settling into a disc. It has lately become clear that a realistic description of star formation and feedback are indispensable ingredients in numerical models if they are to match the observed properties of galaxies  \citep{som99,som03,gov04,gov07,gov10,scann08}.

The formation and evolution of well--ordered systems such as spiral galaxies remains an intriguing process, raising many questions, one of them being the origin and evolution of low surface brightness (LSB) galaxies. Although initially this category was thought to consist predominantly of late--type spirals and dwarf irregular (dIrr) galaxies, we now know that they span the entire range of galaxy mass \citep{bot97}. They are thought to evolve more slowly, as suggested by their low surface brightness, low metallicities and low star formation rates when compared with High Surface Brightness (HSB) galaxies \citep[see][and references therein]{imp97,mih99,rah07}. This also means that LSB discs have consumed less of their gas and are therefore more gas--rich when compared to HSB discs \citep{mcg97,one04}.

An extreme category of LSBs is collectively known as that of `crouching giants' \citep{dis87} or `Malin\,1 cousins' after the giant LSB galaxy Malin\,1, discovered serendipitously by \citet{bot87}. It has one of the largest optical radial extents known for a spiral galaxy, out to a radius of $\sim 100$ kpc \citep{imp89} with an extrapolated disc central surface brightness  of $\mu_0$(V) = 25.5 mag arcsec$^{-2}$. It is one of the most gas--rich galaxies known with an  HI mass of $\sim 6.8 \times 10^{10}$\msun\, and the neutral gas extending out to $\sim 120$ kpc radius \citep{pic97}. The discovery of Malin\,1 prompted searches for more of these giant LSB system resulting in F568--6 or Malin\,2 being found by \citet{bot90}, and 1226$+$0105 by \citet{spr93}. All these `Malin\,1 cousins' were found to be gas rich and have discs characterised by large scale--lengths and low central surface brightnesses as reported by \citet{spr95}, who increased the number of known LSB giant discs by a further eight objects. More recently, using the fact that giant LSB galaxies are rich in neutral gas, \citet{one04} produced a list of 38 additional giant LSB candidates, raising to roughly 55 the number of LSB galaxies with \textsc{H\,i} mass higher than 10$^{10}$\msun. It should be noted, though, that spatially resolved imaging, in \textsc{H\,i} as well as in the optical, is needed to confirm which of these giant LSB candidates are indeed Malin\,1--type objects.

The formation and evolution of LSB giants remains puzzling and several explanations have been put forward. \citet{hof92} claim that rare $3 \sigma$ density fluctuations in a primordial Gaussian random field occurring in voids can lead to objects with the characteristics of a Malin\,1, i.e. a rather normal bulge and a low surface brightness disc. \citet{dal97} postulate that the initial mass and angular momentum of a protogalaxy determine the luminosity and surface brightness of today's systems, high angular momentum protogalaxies leading naturally to giant LSBs.

Rather than their characteristics being due to their initial conditions, some authors like \citet{nog01} suggest that the formation and strength of a bar  in the disc of an HSB galaxy could result in the redistribution of matter, lowering the central brightness of  the object and converting it into a giant LSB galaxy. More recently,  \citet{map08} propose that giant LSBs might be the late stage evolution of a head--on collision, i.e. the late stage of a ring galaxy at $\sim 1.5$\,Gyr after impact.
 
In this paper we present NGC\,765, a galaxy which resembles the giant LSB Malin\,1 but is closer by a factor of 4--5 and hence provides us with an opportunity to study such an extraordinary system in more detail. We present NRAO\footnote{The National Radio Astronomy Observatory is a facility of the National Science Foundation, operated under cooperative agreement by Associated Universities, Inc.} Very Large Array (VLA) D--array archival data of this galaxy. Section~2  provides an overview of archival and literature multiwavelength properties of NGC\,765. In Section~3 we briefly discuss the data reduction for the \textsc{H\,i}, radio continuum, optical, and X-ray data. We present our results in Section~4, where we focus on the morphology and kinematics of the gaseous  disc and provide a brief discussion in Section~5.

\begin{table*}
{\footnotesize
\caption{General information on NGC\,765 and companion galaxies.}
\begin{center}
\begin{tabular}{llllc}
\hline
\hline
 & NGC\,765  & NGC\,765a & UGC\,1453  & Reference\\
\hline
Type  & SAB(rc)bc & \ldots & Irr  & 1\\
RA (J2000)\ &01h 58m 48s & 01h 59m 55.9s~$^3$  & 01h 58m 45.8s & 1\\
DEC (J2000)\ & 24\degr 53\arcmin 32.9\arcsec &  24\degr 58\arcmin 29.9\arcsec ~$^3$ & 24\degr 38\arcmin 34.6\arcsec & 1 \\
Distance & 72 Mpc & 72 Mpc $^3$ & 72 Mpc  & 1\\
V$_{\mathrm hel}$ & 5118 \kms & 4900 \kms~$^3$ & 4897 \kms  & 1\\
R$_{25}$ & 62.6\arcsec & \ldots & $\sim$1.4\arcmin & 1\\
m$\rm _B$ & 13.6 & \ldots & \ldots & 1\\
$\rm L_B$ & $3 \times 10^{10} \lsun $& \ldots  & \ldots & 2 \\
Inclination & 21\degr & \ldots  & 30\degr~$^4$ & 3\\
Position Angle & 251\degr& \ldots & \ldots & 3\\
H\,I flux\ & 38.3 Jy \kms  & 0.53 Jy \kms & 1.54 Jy \kms & 3  \\
H\,I mass & $4.7 \times 10^{10}$ \msun & $6.5 \times 10^{8}$ \msun  & $1.9 \times 10^{9}$ \msun & 3 \\ 
\hline
\hline
\end{tabular}

\end{center}
\label{properties}
}
\hspace{1cm}
\begin{flushleft}
Reference: 1 - Taken from NED; 2 - Derived using M$\rm _{\odot,B}=5.48$ and a distance of 72 Mpc; 3 - This study; 4 - Taken from LEDA.
\end{flushleft}   
\end{table*}

\section{NGC\,765 archive properties}
\label{765 archive properties}

 NGC\,765 (also known as UGC\,1455) is classified as a late-type SAB(rb)bc galaxy \citep{deVau91}, and presents the signature of a central bar.  According to the NASA/IPAC Extragalactic Database (NED) it has a heliocentric recession velocity of 5118 \kms. Throughout this paper we assume a distance of 72 Mpc based on H$_0$=72 \kms Mpc$^{-1}$, which makes $1^{\prime\prime} = 350$\,pc. The galaxy has an extrapolated disc central surface brightness of $\mu$(0) = 22.27 in the $B$--band \citep{dejong96}, or only about 0.5 mag fainter than the \citet{free70} value and therefore brighter than the $\mu (0) \geq$ 23 limit usually adopted to define an LSB \citep{imp97}. The galaxy was included in a sample of (LSB) galaxies by \citet{schombert98}. The detection of nuclear emission of  $\lbrack \rm N\,II \rbrack$, $ \lbrack \rm S\,I  \rbrack $  and   $ \lbrack  \rm O\,I \rbrack $ lines suggests the presence of a  low-luminosity AGN (hereafter LLAGN), which according to \citet{schombert98} are more likely to be found in \textsc{H\,i}--rich, large galaxies. This galaxy and its closest companion, UGC 1453, are thought to be part of  a group of 14  galaxies [GH83]024  \citep{gar93}. In Table~\ref{properties} we summarise some basic information on NGC\,765 and its companion galaxies, stating in advance that NGC\,765a is an uncatalogued  companion discovered in this study. Both NGC\,765a and UGC\,1453 are discussed in more depth in Section~\ref{Results}.

\section{Observations and Data Reduction }

\subsection{H\,I data reduction}
\label{HI data reduction}

The  VLA D--array observations of  NGC\,765 are in the public domain under  project ID AS0625. The galaxy was observed in a set of 3 runs, on  1998 January 17, 18 and 19,  over periods of 5 hours each on source. Both polarisations were recorded. All data reduction was performed using the 31DEC06 version of Classic AIPS. Table \ref{setup} summarises the observational setup.

\begin{table}
{\footnotesize
\caption{Observational setup for the NGC\,765 VLA data.}
\begin{center}
\begin{tabular}{ll}
\hline
\hline 
Parameter &Value  \\
\hline
Object  &  NGC\,765  \\
Instrument  &VLA \\
Configuration & D--array \\
Project ID& AS0625 \\
Primary calibrator & 3C48 \\
Secondary calibrator & 3C48  \\
Central velocity &5189 \kms  \\
FWHM of primary beam &32\arcmin \\
Total bandwidth & 3.125 MHz \\
Number of channels & 64 \\
Channel spacing &0.049 MHz \hspace{0.5cm} (11 \kms) \\
FWHM of synthesised beam & 43.5\arcsec $\times$ 42.3\arcsec \hspace{0.35cm} ($\sim$ 15 kpc) \\
rms noise &0.2 mJy\,beam$^{-1}$ \hspace{0.03cm} (0.065 K) \\
\hline
\hline
\end{tabular}

\end{center}
\label{setup}
}

\end{table}

3C48 was observed interspersed with the target source, in scans lasting 3 minutes. This calibrator was used both as primary calibrator (flux and bandpass) with an estimated flux of 16.50 Jy at 1395.29\,MHz as well as to determine the complex gain (i.e., as secondary calibrator). Antenna 12 was flagged in both polarisations in the first and last runs since visibility amplitudes deviated by approximately 30 percent relative to the average. Flagging and bandpass calibration  were performed with standard tasks.

We deleted the first and last eight velocity channels due to sensitivity loss at the edges of the 3125 kHz wide band  centred at a frequency of 1396.5 MHz. The {\em uv--}data of individual runs were merged into a single data set using the task \textsc{dbcon}.

The data cube was generated using  {\sc imagr}. We employed a \textsc{robust=0} weight in order to obtain the best compromise between resolution and sensitivity. The cube was cleaned to a flux threshold of $2\sigma$. We separated real emission from noise following  the conditional blanking method employed by \cite{wal08}. We convolved the cube to  70 arcsec  resolution, storing in a masking cube regions of emission that were above $2\sigma$ over three consecutive velocity channels. Using the latter cube we then masked our original  43.5 $\times$ 42.3 arcsec cube. The \textit{zero} and \textit{first}  order moment maps (total surface brightness and velocity field), were created with the task {\sc xmom}. Because of the modest angular resolution the channel maps could be cleaned down to the noise and therefore there was no need for residual flux  scaling \citep{wal08}.

NGC\,765 was detected from 5015 \kms\, to 5239 \kms\, in velocity. Its integrated \textsc{H\,i}  spectrum is shown in Figure~\ref{spectrum}. The corresponding channel maps are presented in Appendix~\ref{appendix-a}. In Fig. \ref{765_HI_FIELD} we present an  \textsc{H\,i} column density map of NGC\,765 and its  two nearby companions: UGC\,1453 to the South of NGC\,765, and to the North--East the uncatalogued small companion NGC\, 765a. 

\begin{figure*}

\centering 
\includegraphics[trim= 70  370  70 100, clip, width=12 cm ]{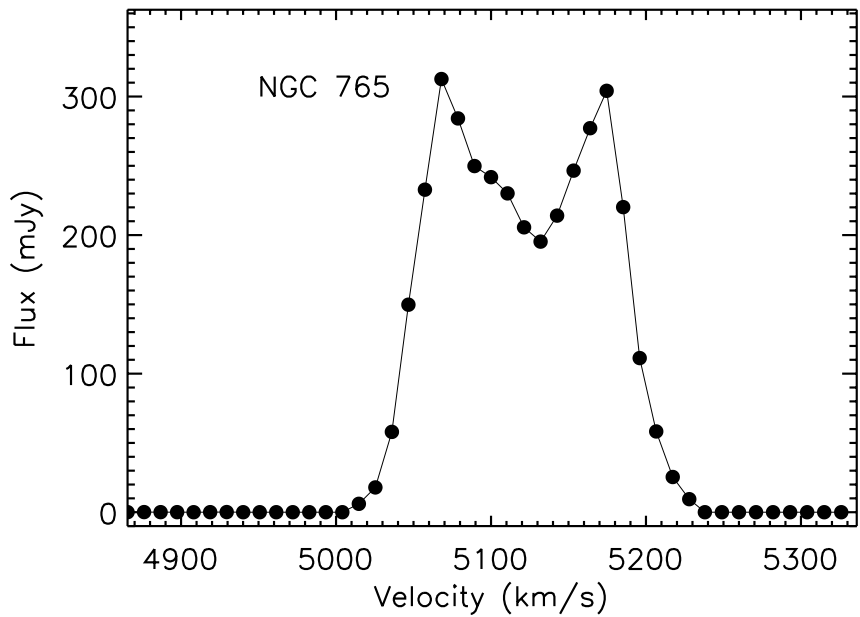}
\caption {Integrated \textsc{H\,i} spectrum of NGC\,765. Note that the integrated flux in each channel map is based on the blanked image cube, hence the flux being at exactly zero for those channels deemed without emission.}

\label{spectrum}
\end{figure*}

\begin{figure*}

\centering 
\includegraphics[trim= 100  150  70 250, clip, angle=-90, width=11 cm ]{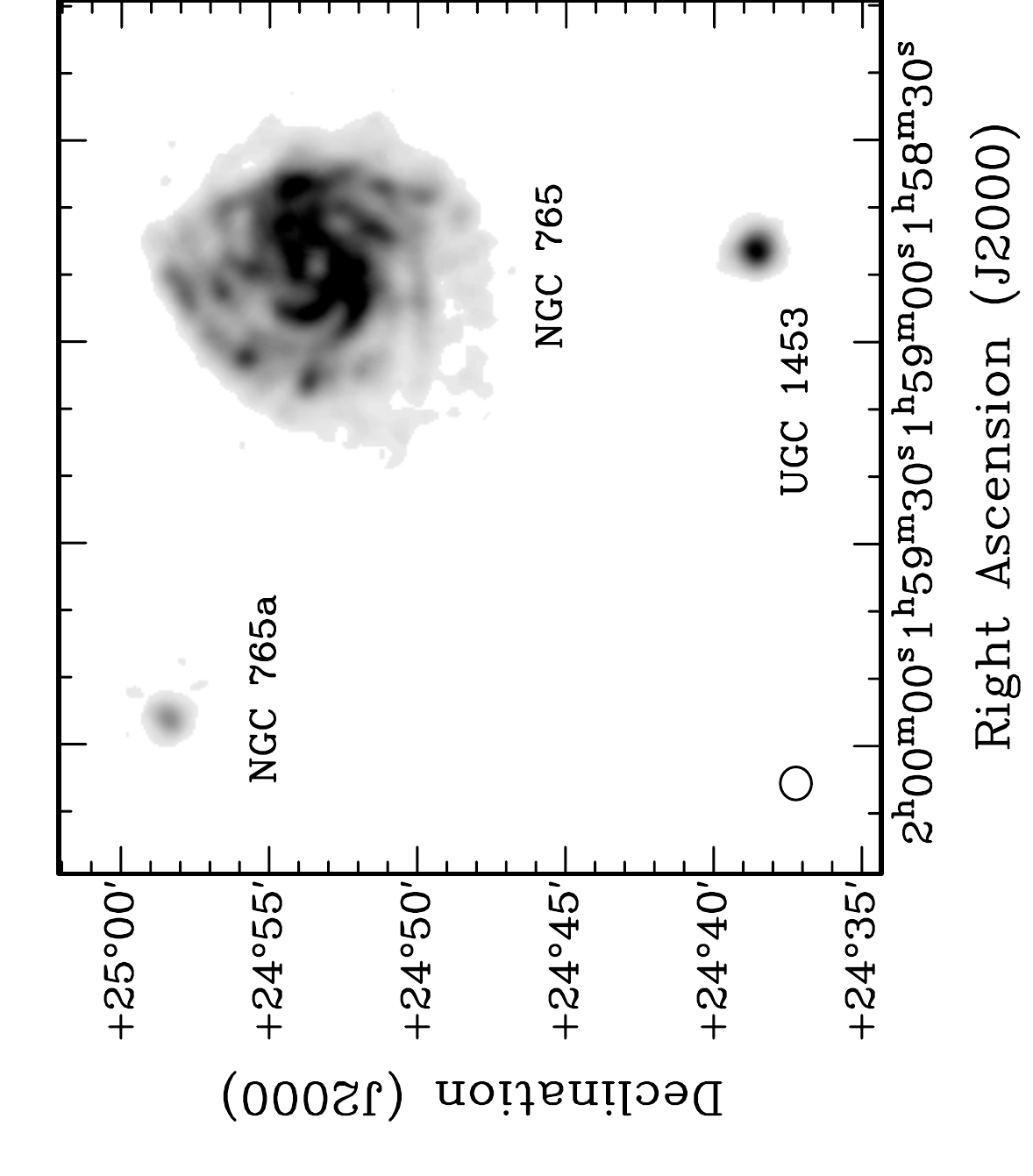}

\caption {Total intensity greyscale \textsc{H\,i} map of NGC\,765. The greyscale ranges from 0 to 225   K  km s$^{-1}$. The resolution is  43 $\times$ 42 arcsec and is indicated by the circle in the bottom left corner of the image. NGC\,765a and UGC\,1453 can be seen to the Northeast and South of the main galaxy, respectively.}

\label{765_HI_FIELD}
\end{figure*}

We  constructed a radio--continuum map using the first five channels of the 48 available, which were thought to be free of line emission. These five channels were employed to perform the continuum subtraction of the above mentioned data cube. The \textit{uv} data of the channels were averaged in frequency using the task \textsc{avspc}, mapped and  {\sc clean}ed with {\sc imagr} in a similar manner as described above. The 1$\sigma$ noise level corresponds to 0.15 mJy. In Fig. \ref{continuum}, we present the continuum emission contours at 1395.39 MHz. 

We detect a source with a flux density of 2.16 mJy (L$\rm _{1.4\, GHz} = 1.3 \times 10^{21}$ W Hz$^{-1}$, for an assumed distance of 72 Mpc), coincident with the optical nucleus of NGC\,765. The NRAO VLA Sky Survey \citep[NVSS;][]{con98} at 1.4 GHz found  no radio continuum emission associated with this  galaxy (Stokes I at a 3 $\sigma$ rms of 1.5 mJy). The Faint Images of the Radio Sky at Twenty--Centimeters (FIRST) VLA--survey \citep{bec95} does not cover this area. No continuum emission was detected for NGC\,765a and UGC\,1453.  We found another 7 radio--continuum sources in the field of NGC\,765. For further details we refer the reader to the material in Appendix~\ref{appendix-b}.

\subsection{Optical data reduction}
\label{optical}

\begin{figure*}

\includegraphics[trim= 10 80 10 50,width=7.0 cm,angle=-90]{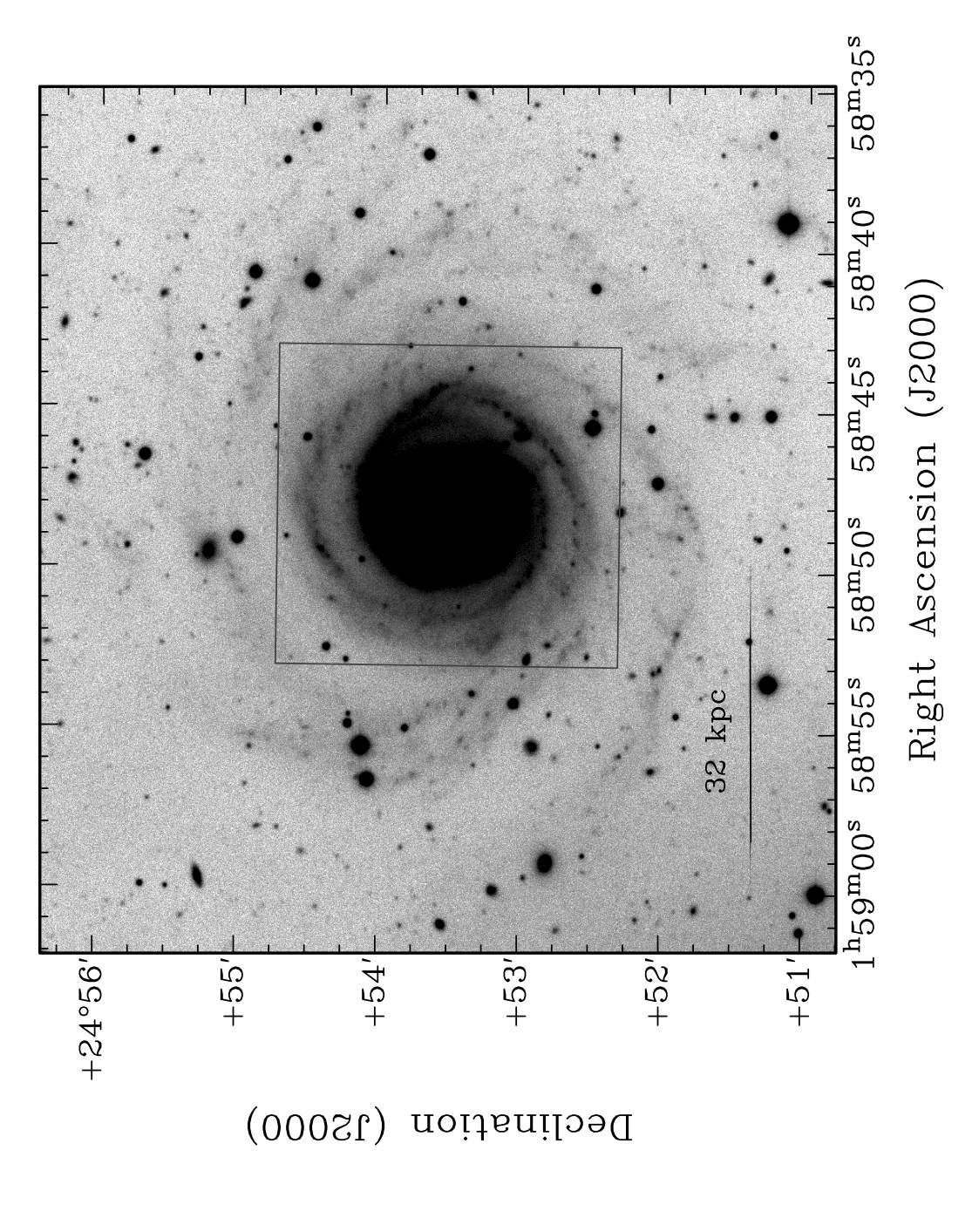}
\includegraphics[trim= 65 -30 -10 100, width=7.4 cm,angle=-90]{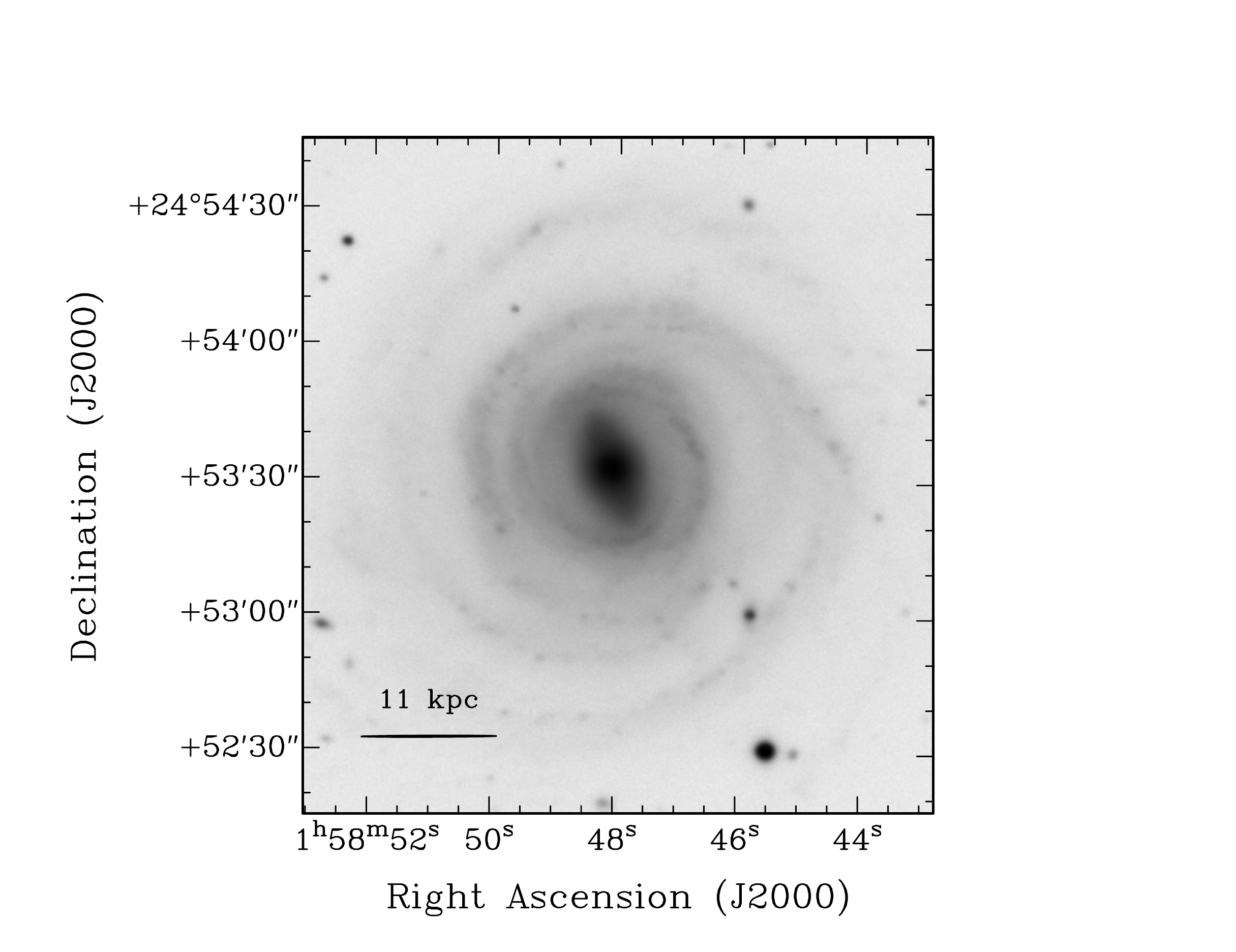}

\caption{Optical image of NGC\,765 acquired with Wide Field Camera  on the INT. On the left, we saturate the central core to bring out the low--luminosity spiral arms. On the right, we zoom in on the central region indicated by the box in the left--hand panel, to expose the central bright core and bar.}
\label{INT}
\end{figure*}

Optical observations of NGC\,765 were acquired with the Isaac Newton Telescope (INT) on 2009 January 22, with the Wide Field Camera (hereafter WFC). The galaxy was observed in three exposures of 600 seconds in the Sloan $R$ filter at a central wavelength of 6240 \AA. Data reduction followed the standard Cambridge Astronomical Survey Unit (CASU) pipeline. In Fig.~\ref{INT} we present the optical INT image of NGC\,765. The PSF is $\sim$ 1.3 arcsec or 0.45 kpc. The galaxy presents an inner bright core and what seems to be a bar oriented NNE--SSW. It is not clear if there is an  associated  stellar ring. Stellar arms, tightly wound around the core, indicate disc rotation in an anti--clockwise direction. Faint traces of at least two arms can be seen as far out as 45 kpc from the nucleus. Both arms seem to bifurcate in the outskirts.  According to \cite{dej94}  the optical diameter measured at  25 mag\,arcsec$^{-2}$ in the $B$--band is 62.6 arcsec (22\,kpc).

We produced an optical surface brightness profile based on elliptical isophote fitting using the projection on the sky as determined by our rotation curve fit (Sect.~\ref{Rotation curve analysis}). Obvious foreground stars were removed and replaced by 2--D interpolation using pixel values of an annulus surrounding the affected area. A 2--D second order spline  was fit to remove the background. Although the background fit produced a remarkably good result, we could see some slight systematic offset due to the bright foreground star towards the south--east corner of the field, potentially affecting the area due South of the  galaxy. We therefore restricted the radial profile, using only the northern half of NGC\,765, employing 3\arcsec\ radial  bins.

\begin{figure*}
\centering 
\includegraphics[trim= 80 350 60 60, clip, width=12.0 cm]{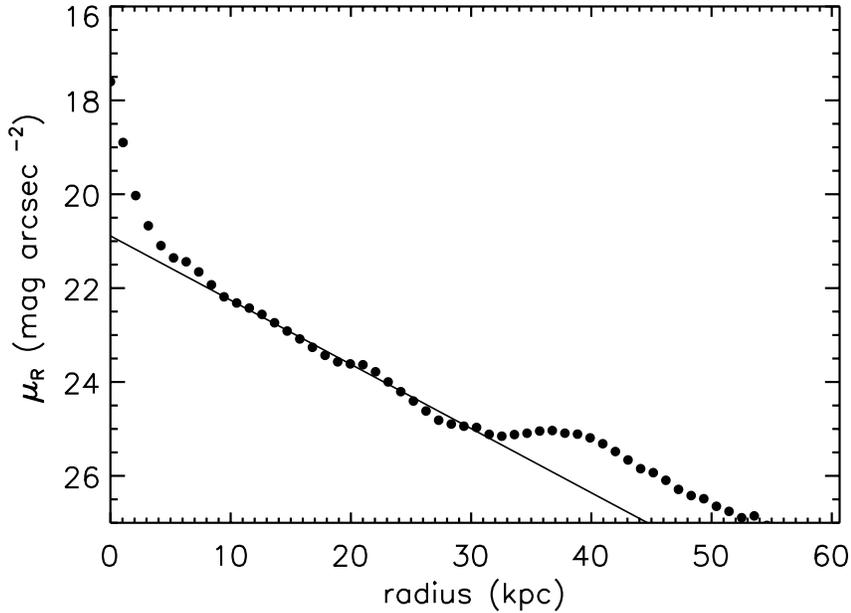}\\

\caption{Optical surface brightness profile of NGC\,765 derived from elliptical isophote fitting on the INT WFC $R$--band image. Only the northern half of the galaxy was used. We also shown the fit to extrapolate the central surface brightness.}
\label{profile}
\end{figure*}

The resulting profile is shown in Fig.~\ref{profile}. The INT data were not calibrated so we bootstrapped our radial profile to the one published by \citet{dej94}. Where our profiles overlap, agreement is excellent. Our data go deeper by almost 2 magnitudes in $R$, down to 26.5\,mag\,arcsec$^{-2}$, allowing us to follow the faint stellar disk out to about 50\,kpc in radius. We derive a scale length for the stellar disc between 5 and 25\,kpc of 7.8\,kpc in good agreement with \citet{dej94} who found values of 8.4\,kpc and 5.7\,kpc in $B$ and  $K_\mathrm{S}$, respectively. The extrapolated central surface brightness at $R$--band is 20.9\,mag\,arcsec$^{-2}$ which, assuming a $B-R = 1.40 $ \citep[taken from][]{dej94} then agrees with their $B$--band value of $\mu_{\mathrm B}(0) = 22.27$. 
Between 30\,kpc and 38\,kpc the profile shows a secondary maximum, beyond which, it declines again at the same rate. This feature can be traced to a faint, broad stellar arm, i.e., this is not an artefact or due to a problem with background subtraction.

\subsection{X--ray data reduction}
\label{Xray data reduction}

\citet{das09} present 0.5 arcsec resolution \emph{Chandra} X--ray observations of a small sample of Malin\,1--type galaxies, including NGC\,765. As they only refer to emission coming from its centre, we decided to retrieve the observations from the archive (project ID 7764)  and independently reduce and analyse them. The data were obtained with the ACIS--S instrument (AXAF CCD Imaging Spectrometer) on 2007 July 11, for a total exposure time of 3.79 ks. Data reduction of the observations was performed using standard procedures included in the \emph{Chandra} Analysis of Observations package (\textsc{ciao} version 4.0). We ran \textit{wavdetect}  with a 3$\sigma$  detection threshold  in order to detect point sources in the NGC\,765 \textsc{H\,i} field. Six sources were detected above 3$\sigma$ and are shown in Appendix~\ref{appendix-c}, Fig.~\ref{xray_figure}.  We confirm emission coincident with the galaxy nucleus at an  X--ray luminosity in the  0.2--10 keV energy band of L$_{\rm X} 1.7 \times 10^{40}$ erg s$^{-1}$, in agreement with \citet{das09}.

\section{Results}
\label{Results}

\subsection{NGC\,765 H\,I   content and morphology}
\label{HIcontent}

 We detect an \textsc{H\,i} integrated flux of 38.3 Jy km s$^{-1}$ which, assuming as usual optically thin emission, corresponds to a total \textsc{H\,i} mass of 4.67$\times 10^{10}$ \msun. This is consistent with the Westerbork Synthesis Radio Telescope (hereafter WSRT) wide-field survey \citep{bra03}, where it was derived to have a total \textsc{H\,i}  flux of  $39.7 \pm 0.66$ Jy \kms (effective beam size of $ 50.3 \times 46.8$ arcmin) and a total \textsc{H\,i} mass of $ 4.6 \times 10^{10}$ M$_{\odot}$. Arecibo observations \citep{shu94}, with a beam of $3.3  \pm 0.1$ arcmin, gave  an integrated \textsc{H\,i}  flux of 11.1 Jy \kms \, which implies an \textsc{H\,i} mass of $ 1.4 \times 10^{10}$ M$_{\odot}$, but these observations were based on a single pointing at the centre and therefore missed a substantial part of the  \textsc{H\,i} belonging to NGC\,765. The total \textsc{H\,i} mass derived for NGC\,765 exceeds by one order of magnitude the median \textsc{H\,i} mass in the LSB survey by \citet{one04}. It also exceeds by a factor of 3 the mean value obtained  by \cite{rob94} for Sab--Sbc HSB spiral galaxies within the RC3-UGC sample.  The \textsc{H\,i} mass to luminosity ratio for NGC\,765 is M$_{\rm HI}$/L$_{\rm B} \sim 1.6$, i.e., of the same order as the median M$_{\rm HI}$/L$_{\rm B} = 2.8$ found for LSBs by \citet{one04}, 
and 5--10 times larger than the median for HSB galaxies \citep{rob94}.

\begin{figure*}

\centering 
 \includegraphics[ width=8.8 cm, trim=0 0 30 0, clip]{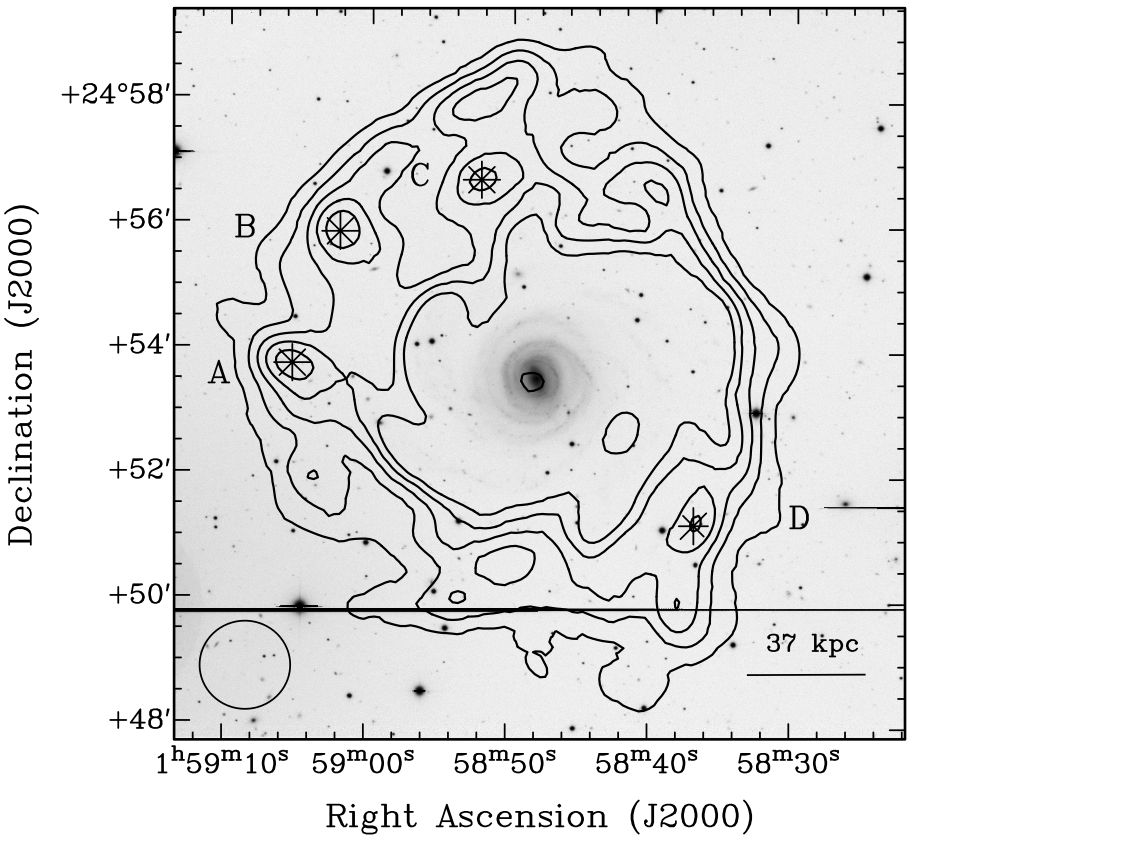}
 \includegraphics[ width=8.2 cm ]{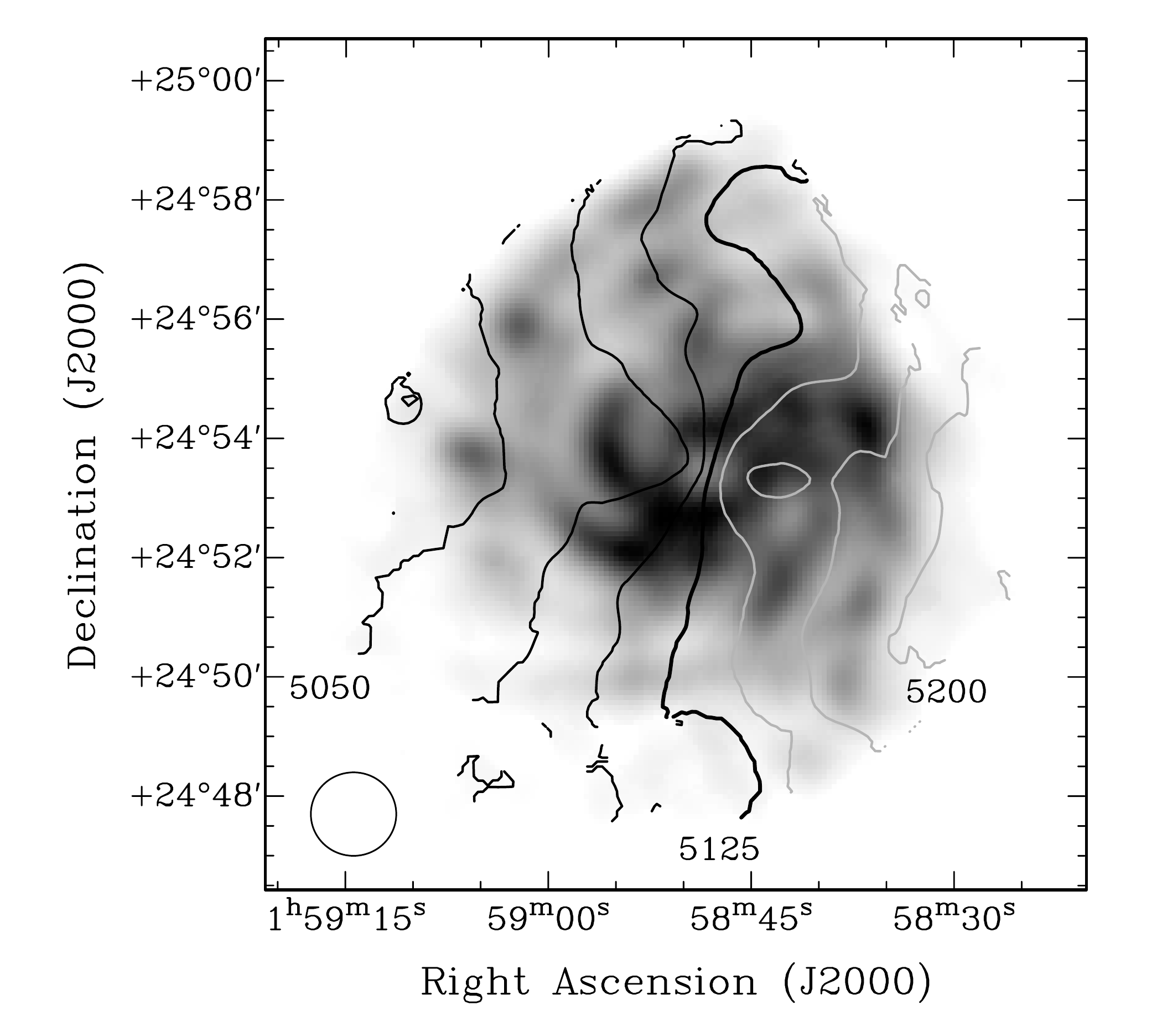}

\caption { Left: \textsc{H\,i}  contours overlaid on the INT optical image. The contours are at a level of 0.5, 1, 1.5, 2 and 2.5 $\times 10^{20}$ atoms cm$^{-1}$. Four compact  \textsc{H\,i} regions are marked with asterisks and labeled A to D (see Sect.~\ref{HIcontent}). The horizontal stripe at a declination of approximately $24^\circ 50^\prime$ is due to bleeding of a bright foreground star along a CCD column (the  original orientation of the CCD was rotated by $90^\circ$). Right: Velocity field  of  NGC\,765  overlaid on an \textsc{H\,i} surface brightness map. Contours are heliocentric velocities ranging from 5050 \kms~to 5200 \kms, in steps of  25 \kms . The systemic velocity of 5125 \kms~is shown as a thick black contour and a few contours are labeled with their velocity in \kms. The West side of the galaxy is receding (light grey contours). In both maps we represent the \textsc{H\,i} beam size in the bottom left corner.} 
\label{765HI&OPT}
\end{figure*}

 NGC\,765 has an extended \textsc{H\,i} distribution with a diameter of 240 kpc.  In Fig.~\ref{765HI&OPT} (left) we present \textsc{H\,i} contours overlaid on the optical image, highlighting  the disproportionately large \textsc{H\,i}  disc compared to the optical. The \textsc{H\,i} structure shows an unresolved  central depression with a size  $\sim$ 50 $\times$ 30 arcsec, better seen in the right--hand panel of Fig. \ref{765HI&OPT} where we present the \textsc{H\,i}  distribution, overplotted with velocity  contours. The inner gas distribution follows the spiral arms. We can identify at least three of them: two to the East and one to the Southwest. The  average \textsc{H\,i} column density out to 50 kpc, i.e., where the \textsc{H\,i} spiral arms dominate, is $5 \times 10^{20}$ atoms cm$^{-2}$. To the East the column density drops beyond that radius to reach a plateau at  $2 \times 10^{20}$ atoms cm$^{-2}$, populated with  \textsc{H\,i} clumps (see below).

\begin{figure}
\centering 
\includegraphics[trim= 80 350 80 80 , width= 8 cm]{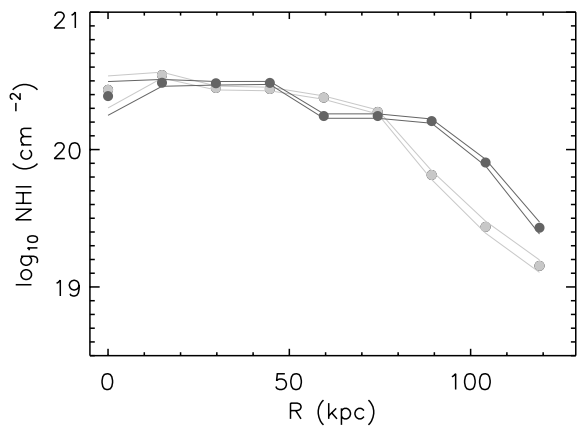}
\caption{Radially averaged surface brightness  profiles of NGC\,765 along the semi--major axis: the western profile (light grey dots) and the eastern profile (black dots). In both profiles the lines bracket the uncertainty in the profile.}
\label{profiles}
\end{figure}

The outer disc of NGC\,765 is asymmetrical or lopsided. This can be appreciated in Fig.~\ref{profiles}, where we present average surface brightness profiles along the major axis: one along the receding side of the galaxy (hereafter West profile) and the other along the approaching side of the galaxy (hereafter East profile). These were created using the ellipse integration  task \textsc{ellint} available in  \textsc{gipsy} \citep{vanhul92}. The profiles are derived from  30\degr \,wide sectors centred around the major axis and binned every 40 arcsec. This process has the advantage of compensating for the loss of signal--to--noise when moving to larger galactic radii by averaging over larger areas, allowing us to recover information all the way out to the periphery of the diffuse disc. In Fig.~\ref{profiles}, we plot column densities (in cm$^{-2}$) on a logarithmic scale as a function of linear radius (in kpc). Each side of the galaxy presents a different signature. Both profiles start out at the same level of $5 \times 10^{20}$ cm$^{-2}$ and show a sudden drop near 50 kpc (2.5 arcmin). The eastern profile settles to a plateau of $2 \times 10^{20}$ cm$^{-2}$ followed by a rapid decline near 100 kpc out to the last measured point at 120 kpc. The western profile declines smoothly from 70 kpc, extending also out to 120 kpc. Both halves together result in a total \textsc{H\,i} extent of 240 kpc, measured at a column density of $2 \times 10^{19}$ atoms cm $^{-2}$.

The low surface density outer  disc of NGC\,765 is populated by several unresolved or marginally resolved compact \textsc{H\,i} regions, notably in the North--East. They are prominent in the integrated \textsc{H\,i} map (Fig.~\ref{765HI&OPT}) as well as in the channel maps (Fig.~\ref{765_CHAN_1}). Spectra taken at their location are indistinguishable from neighbouring areas at the spatial and velocity resolution of the present data, except for their higher surface brightness. Integrated properties of several of these clumps are given in Table~\ref{HICOMPACT_TABLE} (\textsc{H\,i} masses, were derived using task \textsc{ispec}, which retrieves the flux of  a region along individual velocity channels). The compact \textsc{H\,i} regions are labeled in Fig.~\ref{765HI&OPT}.

\begin{figure}

\centering 
\includegraphics[ width=8 cm]{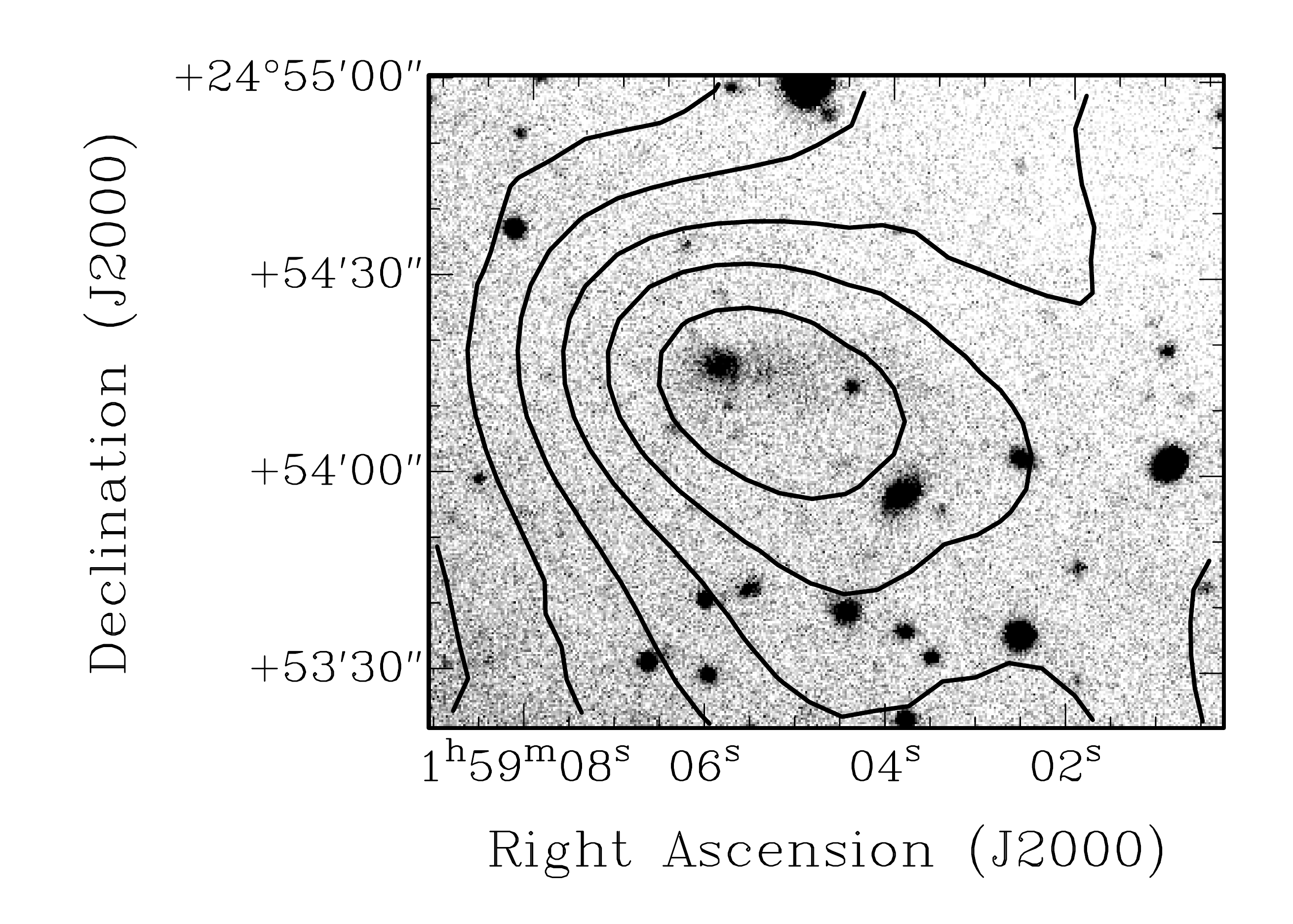}
\includegraphics[ width=8 cm]{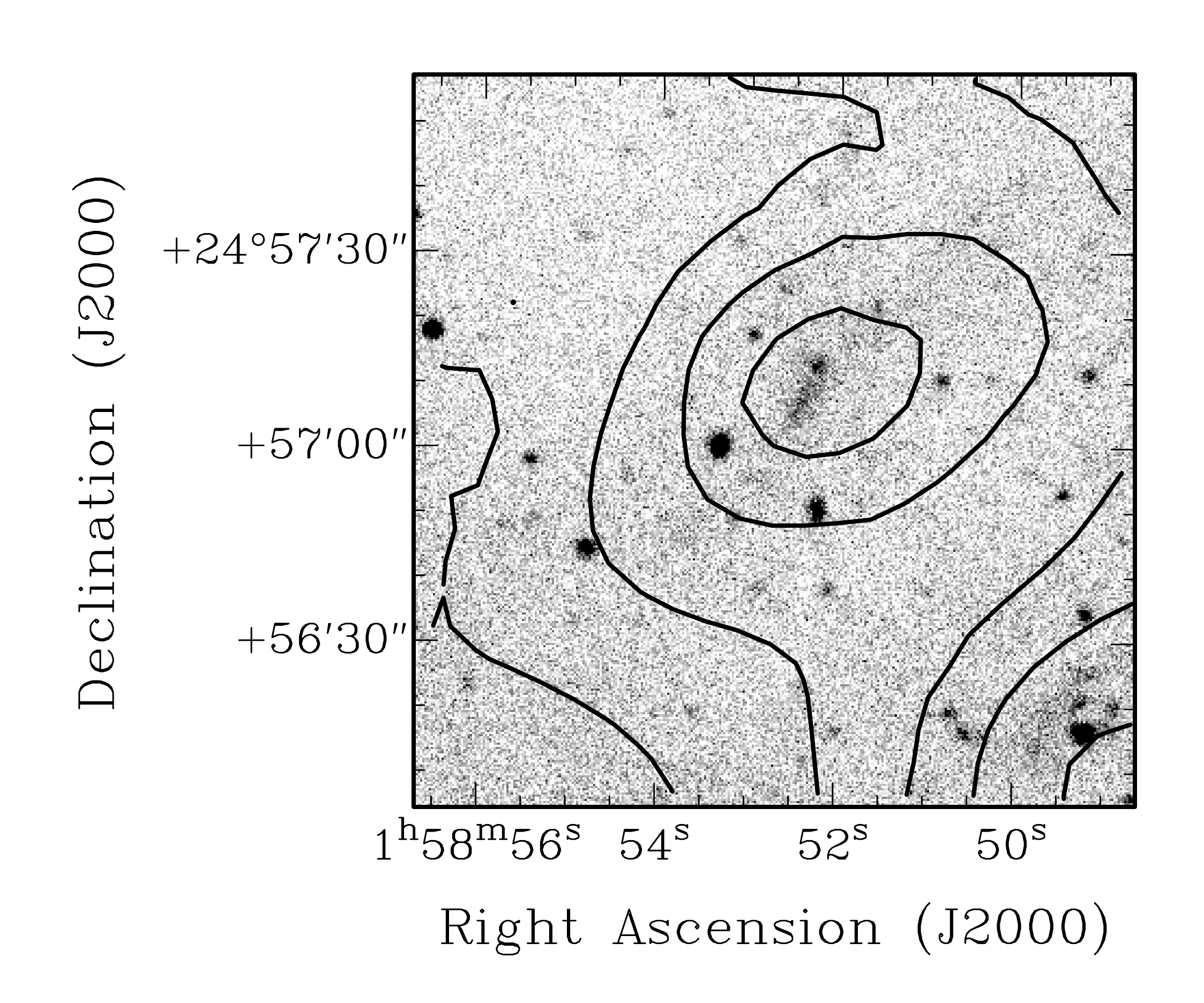}
\caption { Close--up of compact HI regions  A (top) and C (bottom) overlaid on the INT optical image.}
\label{blobs}
\end{figure}

Given that they are at best marginally resolved, their  physical sizes must be less than 10\,kpc; their masses are of order $10^8$ \msun. Together they represent $\sim$ 10 per cent of the total \textsc{H\,i} mass of NGC\,765. Their properties are comparable to those of the companion galaxies detected in our field:  NGC\,765a and UGC\,1453  (see Sect.~\ref{companions}) and fall within the mass interval for \textsc{H\,i}--rich companions found by \cite{pis03}. The  \textsc{H\,i} masses of the compact \textsc{H\,i}  regions are, however an order of magnitude higher than the range quoted by  \cite{wil01} for  \textsc{H\,i} clumps in  discs. In addition, regions A and C present optical counterparts (Fig.~\ref{blobs}). In both of these, the stellar counterparts and the \textsc{H\,i} gas are aligned but interestingly do not follow the spiral arms. Neither of the remaining compact \textsc{H\,i} regions coincides with an optical, radio--continuum or X--ray counterpart.

\begin{table}
\caption{Basic properties of \textsc{H\,i} compact regions}
\centering
\begin{tabular}{ccccc}
\hline
\hline 
Name  & $\alpha$ & $\delta$ & Flux  & Mass \\
& J(2000) & J(2000) & Jy km s$^{-1}$ & 10$^8$ \msun  \\
\hline
A & $01^h\, 59^m\, 04.6^s$    & 24\degr~53\arcmin~39\arcsec  & 2.1  & 10.2   \\
B & $01^h\, 59^m\, 01.5^s$    & 24\degr~55\arcmin~49\arcsec  & 2.2  & 10.5 \\
C & $01^h\, 58^m\, 51.6^s$    & 24\degr~56\arcmin~43\arcsec  & 2.3  &9.2   \\
D & $01^h\, 58^m\, 38.1^s$    & 24\degr~50\arcmin~05\arcsec  & 1.4  & 6.3 \\
\hline
\hline
\end{tabular}
\label{HICOMPACT_TABLE}
\hspace{1cm}
\begin{flushleft}
Notes: Column 1: Source name; Column 2 \& 3:  Right ascension and declination; Column 4: Integrated flux; Column 5: \textsc{H\,i} mass. 
\end{flushleft}   
\end{table}

\subsection{Velocity field and kinematics}
\label{Rotation curve analysis}

The velocity field presented in Fig.~\ref{765HI&OPT} (right) shows steeply rising rotation in the inner part (up to $\sim$ 40 arcsec in radius) that is basically unresolved by our beam and a  flat rotation curve onwards. Between 2.5  $\leq r  \leq$ 3.0 arcmin there seems to be a kink in the isovelocity contours, best seen on the North side. The position--velocity (PV)--diagrammes shown in Fig. \ref{pv_diagrammes} are taken along the major (PA$=256$\degr) and minor axis (PA$=166$\degr) of the galaxy (see below). We see that both Western and Eastern sides are fairly symmetric within the inner 3 arcmin, with velocities rising  quite steeply to a flat level of rotation after just 40 arcsec, and a  jump in velocity at $\sim$ 3 arcmin. The ellipse visible in the minor axis diagramme (Fig.~\ref{pv_diagrammes}, right) is a consequence of the presence of a bar, causing streaming motions in the central part, and our $\sim$ 43 arcsec beam: emission from both the receding and approaching sides enters the beam simultaneously.

\begin{figure*}
\centering 
\includegraphics[angle=-90,trim= 0 50 0 0,clip ,width= 7 cm]{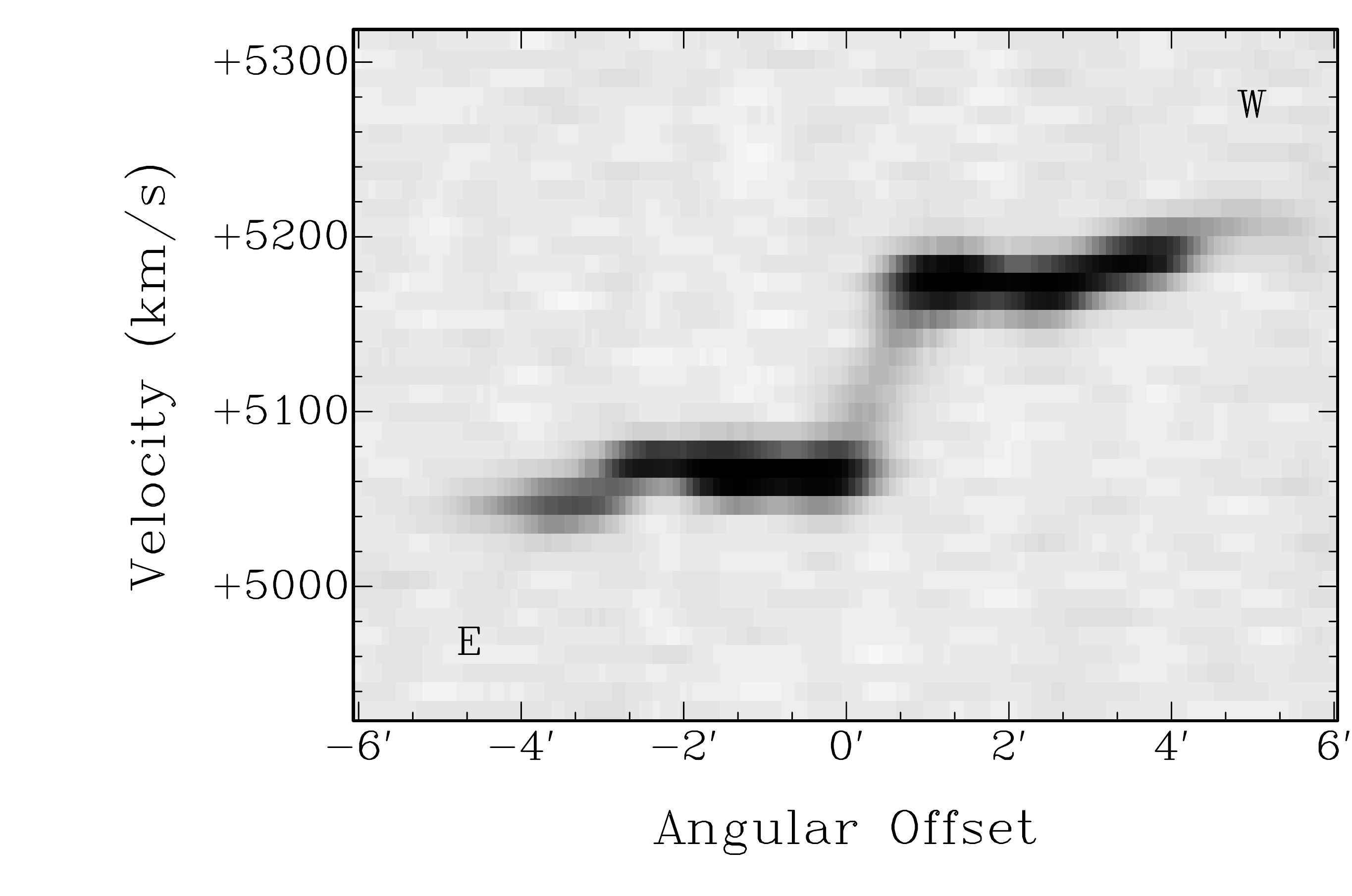}
\includegraphics[angle=-90 ,trim= 0 50 0 0,clip,width= 7 cm]{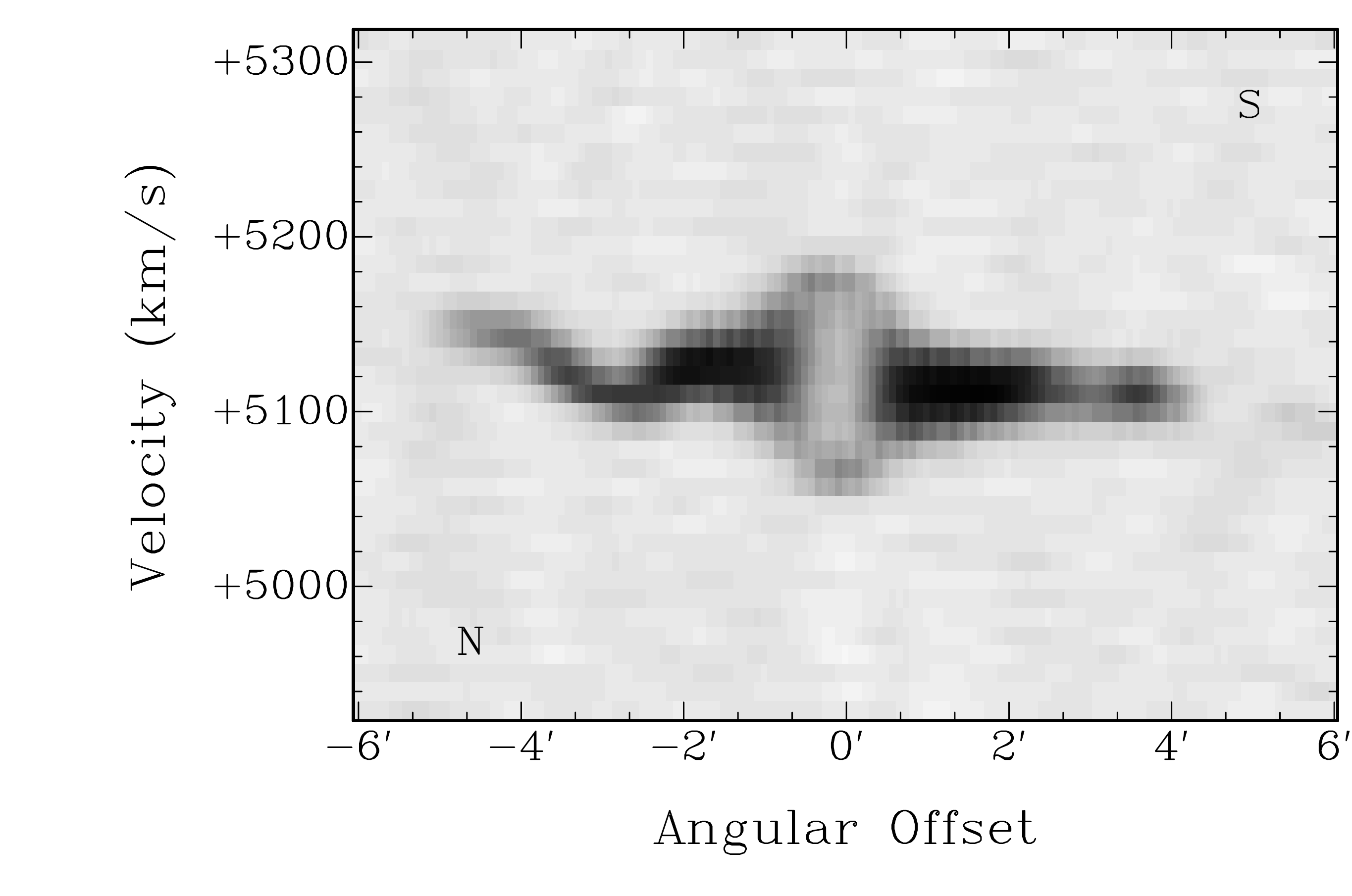}

\caption{ Position-velocity map along the kinematical major axis of NGC\,765  at a PA of 256\degr\ (left panel), showing the rotation of the galaxy. In the right--hand panel we show the position--velocity diagramme taken along the minor axis (PA of 166\degr).}
\label{pv_diagrammes}
\end{figure*}

We used the task \textsc{rotcur}  in \textsc{gipsy} \citep{beg87} to model the velocity field of NGC\,765. \textsc{rotcur} treats the observed velocity field as a series of tilted rings. As already mentioned by \citet{beg87}, this task becomes increasingly unreliable for low inclination (face--on) galaxies. Given the low inclination of NGC\,765 our results should therefore be treated with some caution. Having said that, NGC\,765 provides a unique opportunity to probe its dynamical mass and hence the contribution from the Dark Matter halo out to an unprecedented distance.

It should also be noted that beam smearing will affect the derived rotation curve, mostly in the inner part. This is nicely illustrated in the case of Malin\,1 by \citet{san07} who reanalysed the VLA data obtained by \citet{pic97}, taking beam smearing fully into account. In the case of NGC\,765, with 12 independent beams across the galaxy, beam smearing will affect the derived rotation in the innermost region only and has no impact on our conclusions.
 
 We used concentric rings with a width of 40 arcsec,  matching the resolution of the integrated \textsc{H\,i} map. We fitted for each ring the systemic velocity, rotation velocity, position angle, inclination and central position. Starting values of the systemic velocity and centre position were taken from the optical properties presented in Table~\ref{properties}. The position angle of 270\degr\  was chosen after visual inspection of the data cube and velocity field. The inclination of NGC\,765 in NED is listed as 0\degr. The position velocity diagramme along the major axis shows a clear velocity gradient with a peak--to--peak range of about 150 km s$^{-1}$, arguing against NGC\,765 being nearly face--on. Neither the optical map nor the integrated \textsc{H\,i}  map can be used to derive a reliable inclination based on the minor to major axis ratio. The central region of the optical image is dominated by the bar while the outer disc is of too low surface brightness; the \textsc{H\,i} map is clearly lopsided. We therefore decided to use as initial value 45\degr . After two iterations a systemic velocity of $\rm V_{sys} = 5125$ km s$^{-1}$ was obtained, consistent with the value retrieved from NED of 5118 \kms. 
 
\begin{figure}
\centering 
\includegraphics[trim= 80 320 320 20 , width= 6.8 cm]{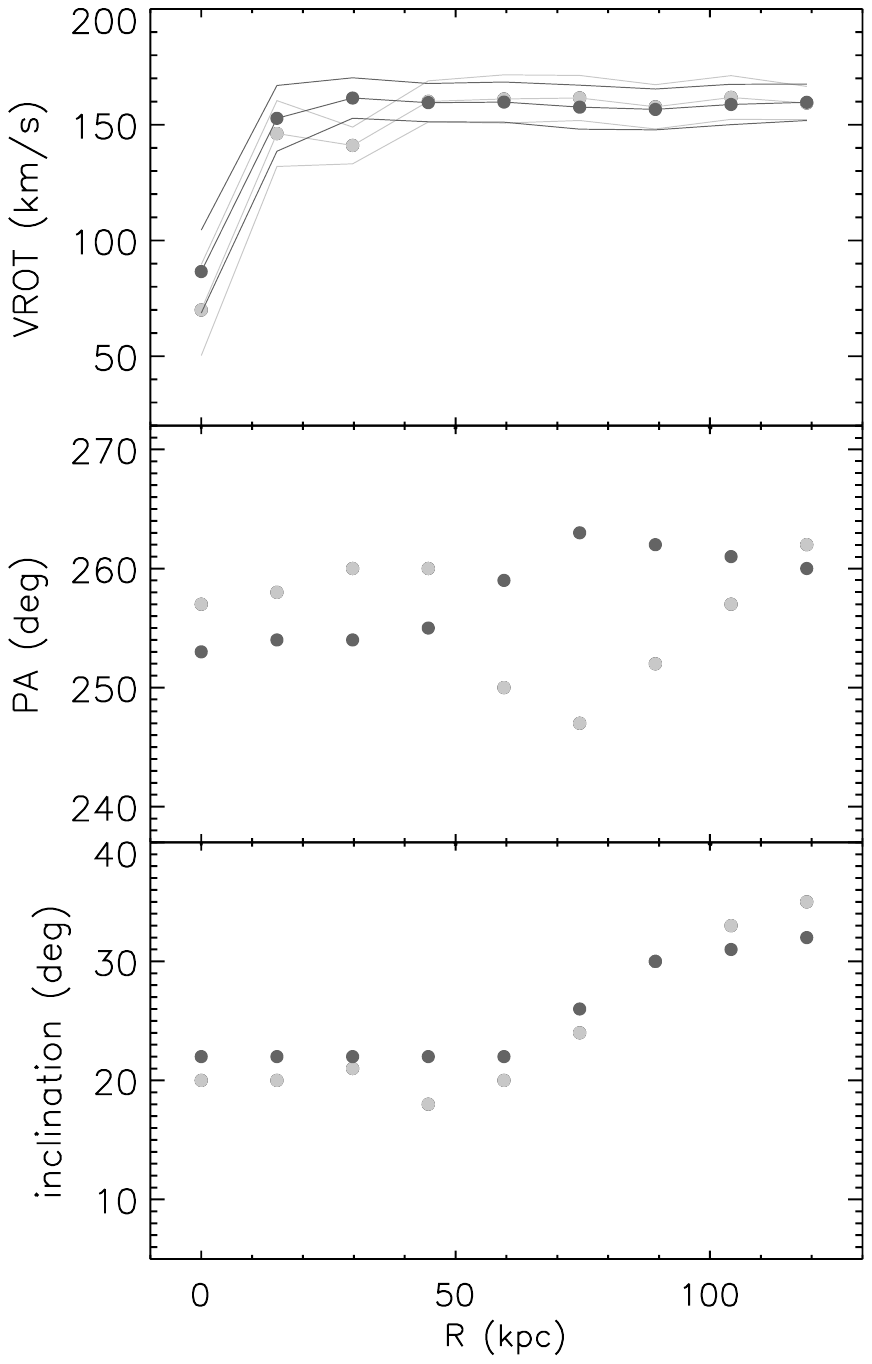}
\caption{Rotation curves for individual halves of NGC\,765 (top panel) with respective values of position angle (middle panel) and inclination (bottom panel) as a function of radial distance in kpc. The black dots represent the kinematically approaching (East) side of the galaxy whereas the light grey dots correspond to the receding (West) side. In the top panel, thin light and dark lines represent the uncertainties.}
\label{rotcur1}
\end{figure}

 We then proceeded to constrain the remaining free parameters. The major axis PV--diagramme shows a jump to slightly higher velocities around 2.5 arcmin radius, near where the \textsc{H\,i} column density drops to a plateau of $2 \times 10^{20}$ atoms cm$^{-2}$, and the galaxy takes on a decidedly lopsided look. Galaxy rotation curves are observed to be flat out to the last measured point and we therefore concluded that the outer \textsc{H\,i} disc must be slightly warped. Hence we fixed the rotation curve to a flat 160 \kms\, beyond $\sim$ 100 arcsec, allowing for a smooth variation in position angle and inclination. We show in Fig.~\ref{rotcur1} (middle and bottom panels) the \textsc{rotcur} results for fits to each side of the galaxy separately. The inclination varies from a constant 21\degr\ in the inner, high \textsc{H\,i} surface brightness part to around 30\degr\, in the outskirts. 
  
The dynamical mass of NGC\,765 was derived using the thin disc approximation \citep{moor82}:
\begin{equation}
 \rm M_{\rm dyn} = 0.76 \frac{ \rm R_{HI}V^2}{\rm G} 
\end{equation}
where R$_{\rm HI}$ is the radius of the last measured point of the rotation curve and V is the rotation velocity respectively. For NGC\,765, at a radius of 320 arcsec or 112 kpc and a velocity of 160 \kms, we derive a dynamical mass of 5.1$\times 10^{11}$ \msun.  

\subsection{Companions }
\label{companions}

\begin{figure*}
\centering 
\mbox{
\includegraphics[ width=8.6cm,trim = 30 20  -90 50 ,angle=-90]{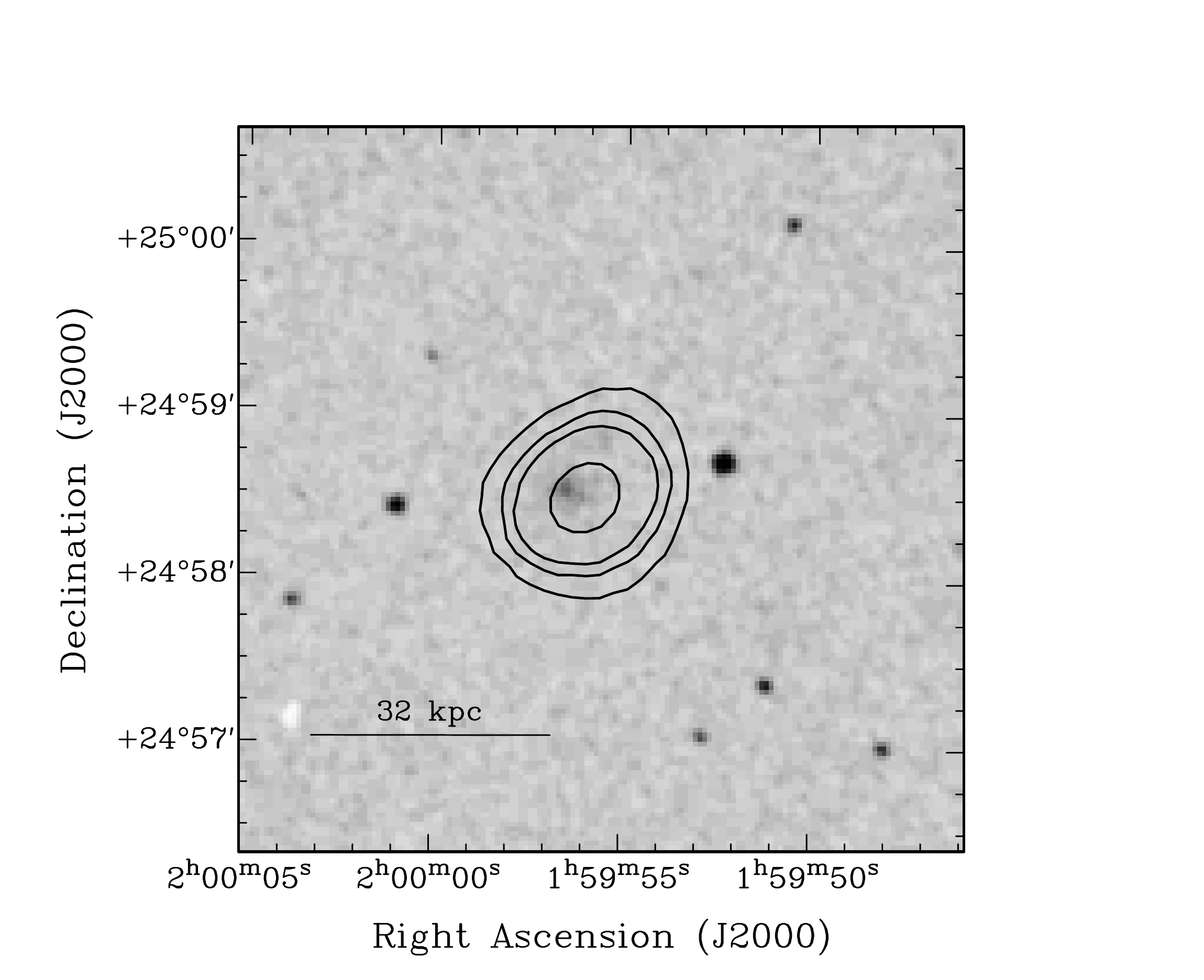}
 \includegraphics[ width=7.4cm, trim = 65 -10 10 100,angle=-90]{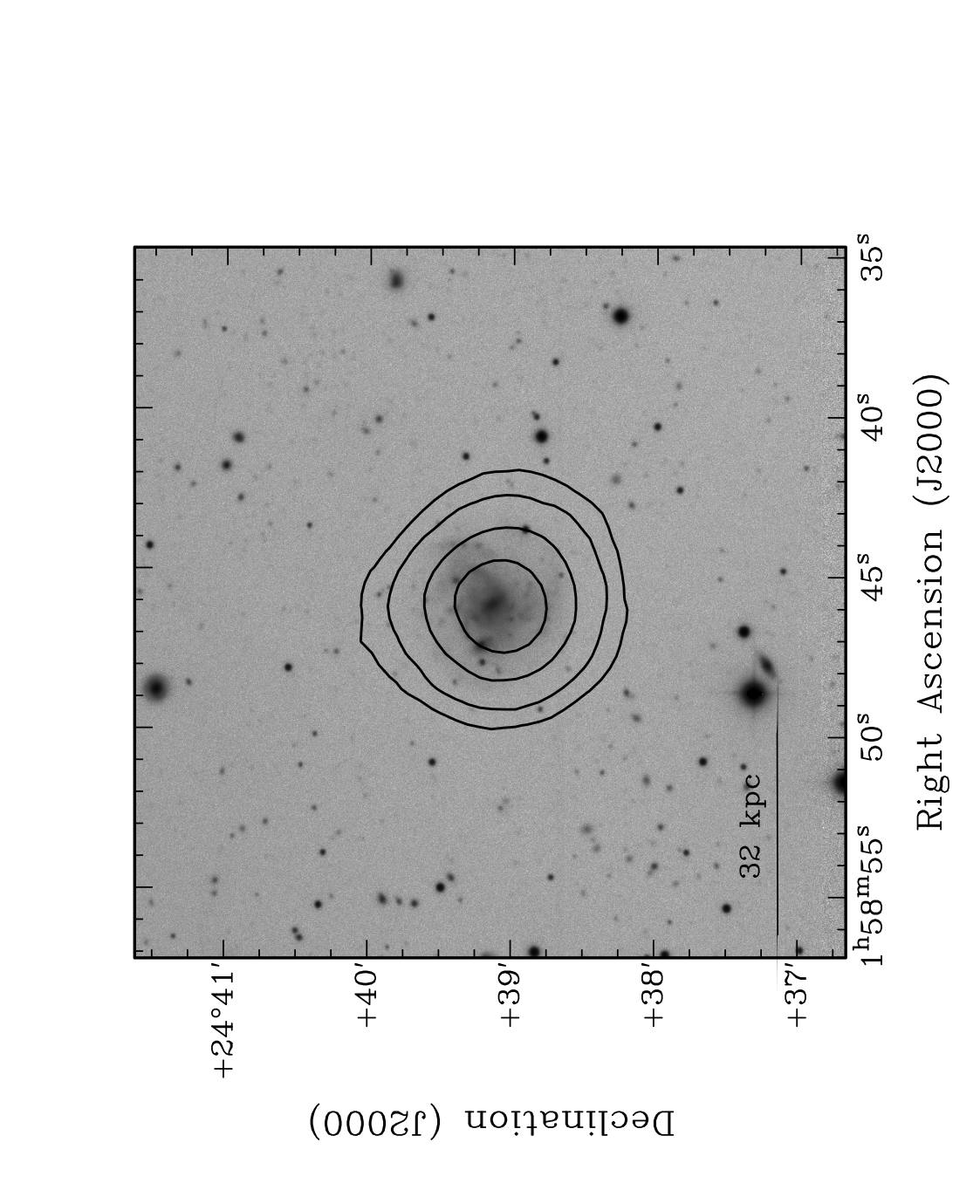}
 }
\caption {The  left--hand panel  presents  \textsc{H\,i}  contours of NGC\,765a  overlaid on a DSS optical image. The contours are 0.5, 0.8, 1 and 1.5 $\times 10^{20}$ atoms cm$^{-2}$. The right--hand panel shows \textsc{H\,i}  contours of UGC\,1453  overlaid on an INT optical image. The contours are at  column densities of 0.5, 1, 2 and 3 $\times 10^{20}$ atoms cm$^{-2}$.}
\label{765comp}
\end{figure*}

  Fig.~\ref{765_HI_FIELD} includes two companions of NGC\,765 detected in \textsc{H\,i}:  NGC\,765a  and UGC\,1453. Both objects are seen across nine consecutive velocity channels towards the low-velocity edge of our bandwidth (4860 -- 4950 \kms). In Fig. \ref{765comp} (left) we see NGC\,765a as \textsc{H\,i} contours overlaid on a DSS optical image. The positional accuracy of the DSS images are typically $1^{\prime\prime}$. A faint optical counterpart coincident with the \textsc{H\,i} peak can be seen. In the case of UGC\,1453, \textsc{H\,i} contours are overlaid on the INT optical image (Fig.~\ref{765comp}, right). It is possible to see faint flocculent, arm--like structure, typical of an irregular galaxy. 
  
 NGC\,765a is a newly detected \textsc{H\,i} source.  It is marginally resolved. A gauss fit on the integrated \textsc{H\,i} map gives a deconvolved size of $13.8  \times 8.5$\,kpc for the assumed distance of 72 Mpc, at a position angle of 134\degr. We used {\sc aips}  task {\sc ispec} to derive a flux of 0.53 Jy km s$^{-1}$, which translates to an \textsc{H\,i} mass of $6.5 \times 10^8 $ \msun. The projected angular distance to NGC\,765  is 16.2 arcmin or 338 kpc and the radial velocity difference is 225 \kms.

Whereas NGC\,765a is uncatalogued, UGC\,1453 is a known 17 mag irregular galaxy at a measured velocity of 4897 \kms (NED). It has as an optical size of 1.5 $\times$ 1.3 arcmin.  Like NGC\,765a, it is marginally resolved with our beam and we derive an intrinsic diameter for the bulk of the \textsc{H\,i} of 17\,kpc. We derive an integrated flux of 1.54 Jy \kms\ and an \textsc{H\,i} mass of 1.9 $ \times 10^9 $ \msun. The mass  falls within values typical for  Sm or Im galaxies \citep{rob94}. Because the detection falls at the  low-velocity edge of the  band, we might be missing some of the integrated flux, given the 2.66 Jy \kms \, value reported by  \citet{gio86} and \citet{sch90} based on Arecibo neutral hydrogen observations. \citet{springob05} list 2.57 Jy \kms. This means that we might also be underestimating its \textsc{H\,i} extent. We derive a projected angular distance from NGC\,765 of 14.8 arcmin or 310 kpc and a radial velocity difference of  221 \kms.

\subsection{Nuclear  activity}

NGC\,765 shows signatures of a bar, a structure that is capable of transporting gas from the inner regions of a galaxy to its nucleus. The presence of nuclear emission lines of $\lbrack \rm N\,II \rbrack, \lbrack \rm S\,I \rbrack $  and  $ \lbrack \rm O\,I  \rbrack $ and the detection of a compact low--luminosity radio--continuum source (Sect.~\ref{HI data reduction}), strongly suggests the presence of an LLAGN. This is further strengthened by the detection of the central X--ray source (Sect.~\ref{Xray data reduction}; Appendix~\ref{appendix-c}; \citealt{das09}). The X--ray luminosity (L$\rm _{X}$ $\approx$ 1.7 $\times$ 10$^{40}$ erg s$^{-1}$) and radio--continuum luminosity (L$\rm _{1.4\,GHz} = 1.3 \times 10^{21}$ W Hz$^{-1}$) are typical of low-luminosity AGN \citep{Ho01,des09}. 

The detection of AGN activity (in radio and X--rays) is quite common, with \citet{des09} quoting 20--25\% for late-type (Hubble types Scd--Sm) galaxies. \citet{fil06} found the same fraction for a sample of  bright local galaxies taken from the Palomar survey of bright galaxies \citep{ho97}. As mentioned by \citet{das09}, very little is known about nuclear activity of LSBs, although those LSBs that have well--developed bulges, like NGC\,765, seem to follow the same evolutionary path as bulge--dominated HSBs.

\section{Discussion}
\label{Discussion}

 \subsection{The extent and morphology of the H\,I  disc}

The inner region (2--4 arcmin or 40--80\,kpc diameter) of NGC\,765 appears to have the normal morphology  of a spiral galaxy, as shown in Fig.~\ref{765HI&OPT}.  It has a central bar as do 70\% of  disc galaxies \citep[e.g.][]{erw05}. The neutral gas distribution, as seen in in Figs.\,\ref{765_HI_FIELD} and \ref{765HI&OPT},  and as described in Sect.\,\ref{HIcontent}, reveals an \textsc{H\,i}--poor central region, which is common in cases of a bar. The \textsc{H\,i} spiral arms trace well their optical counterparts. 

Based on its central surface brightness it is classified as an LSB, its $\mu_{\mathrm B}(0)$ of 22.27\,mag\,arcsec$^{-2}$ falling below the range of $21.65 \pm 0.3$\,mag\,arcsec$^{-2}$ of HSB galaxies \citep{free70}. This is further confirmed by its large 7.8\,kpc scale length for the exponential disk \citep[e.g.,][]{spr95}. NGC\,765 is unique, though, for the linear size of its \textsc{H\,i} disc. The fact that we can trace the neutral hydrogen out to 240 kpc in diameter, makes this  the largest \textsc{H\,i}  disc ever measured in a local spiral galaxy, on a par with the more distant Malin\,1. The total \textsc{H\,i} content of $4.7 \times 10^{10}$\,M$_\odot$ equally puts it in the category of giant LSBs \citep{one04}, as does the resulting 
M$_{\rm HI}$/L$_{\rm B} \sim 1.6$.

There is a clear difference though when we compare NGC\,765 with some of the other well known Malin\,1 cousins in the sense that the radius of its  \textsc{H\,i} disk compared to optical scale length is considerably larger. This is illustrated in Table~\ref{sizes} where we compare NGC\,765 with the  \textsc{H\,i} radius--to--optical scale length of the four  giant LSBs observed by \citet{pic97}.

\begin{table}
{\footnotesize
\caption{ R$_\mathrm{HI}$ to h ratios of some galaxies from the literature and NGC\,765}
\centering
\begin{tabular}{lrrr}
\hline
\hline
  Object  & R$_\mathrm{HI}$ & h\ \ \ \  & R$_\mathrm{HI}$/h \\
  		& (kpc)	&	(kpc) & \\
\hline
Malin\,1			& 100	&  82.0	& 1.2 \\
F568--6 (Malin\,2)	&   90  	& 18.3 	& 4.9  \\
UGC\,6614		&   50	& 13.6 	& 3.7 \\
NGC\,7589		&   39 	& 12.6 	& 3.1 \\
NGC 765			& 120 	&   7.8 	& 15  \\

\hline
\hline
\end{tabular}
\begin{flushleft}
Note: optical scale lengths were determined on $R$--band images except for Malin\,1, where the $V$--band was used.
\end{flushleft}   

\label{sizes}
}
\hspace{1cm}
\end{table}

NGC\,765 can be divided into two regimes, inner and outer disk, with the transition at $\sim 50$\,kpc. At that radius the optical disk has become too faint to be traced any further, we find a jump in velocity in the major axis position--velocity diagramme which we interpret as being due to a modest warping of the disk beyond 50\,kpc, and at this same radius we find that the \textsc{H\,i} column density drops from a constant level of $5 \times 10^{20}$\,cm$^{-2}$ to a plateau of $2 \times 10^{20}$\,cm$^{-2}$.

The outer \textsc{H\,i}  disc is asymmetrical, is at a lower column density,  and is  populated by compact \textsc{H\,i} regions. These regions typically measure $< 10$\,kpc and have \textsc{H\,i} masses of 10$^{8-9}$ \msun \, (of the same order as those of dIrr galaxies). Discrete \textsc{H\,i} clumps lacking an optical counterpart have been found around galaxies, typically within 50\,kpc and 60\,km\,s$^{-1}$ \citep{wil01}, and represent perhaps the fossil evidence of the process of galaxy formation.

Hierarchical galaxy formation predicts the assembly of galaxies via a merger of smaller units. Indeed, there is  observational evidence for major or minor mergers, in the form of interacting galaxies, accretion of dwarfs, \textsc{H\,i} clumps, or in the case of  cold accretion, filaments or streams of material, as highlighted by \citet{san08} and \citet{van05}. The stellar counterparts  present within the compact \textsc{H\,i} regions A and C as shown in Fig.~\ref{blobs} lend evidence that they could be fossil remnants of past mergers in a process similar to the Sagittarius (Sgr)  dwarf galaxy discovered by \citet{iba94}, a tidal stream that is observed near the Milky Way \citep[][and references therein]{mar01}. Simulations performed by \citet{iba98} show that the Sagittarius dwarf with a mass of $\sim 10^9$\msun, would survive several encounters with the Milky Way and produce distortions in the \textsc{H\,i}  disc of the Galaxy. In addition, simulations by \citet{pen06} also point to the fact that an extended outer disc ($\sim$ 200 kpc) of an M\,31 size galaxy could be created via the merger of a dwarf companion with a mass of $\sim 10^9 - 10^{10}$\msun.    

Based on this, it is fair to speculate that over its lifetime, NGC\,765 has accreted  \textsc{H\,i} through a series of merger events with dwarf galaxies with \textsc{H\,i} masses of the order 10$^{8-9}$ \msun, enlarging the size of the  disc and acting as a reservoir of \textsc{H\,i} fuel for future star formation   \citep{pis03,san08}.   As pointed out by the referee, an additional argument in favour of this scenario is that  density enhancements  in the disc of the size of the \textsc{H\,i} clumps  observed here, would loose their identity  due to shear within less than 1 Gyr, unless they are fairly massive and hence accreted companion galaxies.

In support of this, NGC\,765 belongs to a galaxy group,  [GH83]024  \citep{gar93} and the field contains two companions within $\sim$340 kpc,  which have \textsc{H\,i} masses and sizes similar to the compact \textsc{H\,i} regions present in the outer  disc of NGC\,765. The accretion scenario would also fit in with the kinematical and morphological lopsidedness of the \textsc{H\,i}  disc. In their study of the spiral galaxy NGC\,6946, \citet{boo08} interpreted  lopsidedness, signatures of  sharp edges of the \textsc{H\,i}  disc and wiggles in the velocity field as possible indicators of accretion. All these signs are  present in NGC\,765. Besides, if gas is indeed being accreted, it does not necessarily start out in an orbit co--planar with the inner  disc, which could explain why gas in the outer parts resides in a plane with a different orientation. Having said that, the fact that NGC\,765 belongs to a group increases the probability of it having suffered tidal interactions which are able to to produce lopsidedness or excite an oscillation in the z--direction.

 \citet{san08} report that  25\% of galaxies with observational evidence of  tidal interaction have clumps of gas with masses of order $10^{8-9}$\msun, which corresponds to the \textsc{H\,i} masses of the compact \textsc{H\,i}  regions of the outer  disc of NGC\,765. \citet{pis03} show evidence  that these companions are statistically in circular orbits,  their masses are low, and therefore will not strongly affect  the host's morphology.  This is also corroborated by \citet{bou05}, who investigate  via N--body simulations that  mergers with mass ratios in the range from 20:1 to 7:1 will trigger significant lopsidedness in their hosts on time scales of at least a few Gyrs. 
 In our case,  we estimate that NGC\,765a and UGC\,1453 could have  been gravitationally bound to NGC\,765 over the past $\sim 9$ Gyr. If accreted, these would represent between 1 to 4 per cent of its \textsc{H\,i} mass. 
 
 There are two problems, though, with this scenario. One is that the amount of \textsc{H\,i} in NGC\,765 is considerable and that a large number of minor mergers would be required. The other problem is that the period of rotation in the outer disk is of order 4--4.5\,Gyr which makes for a very long relaxation time. So perhaps we are dealing with nature rather than nurture, the structure of NGC\,765 being defined more by the initial conditions of its formation \citep{dal97,hof92} and less by its merger history.

\subsection{Surface brightness profiles and SFR}

What remains intriguing about NGC\,765 is the disparity between the optical and \textsc{H\,i}  discs. With such large amounts of \textsc{H\,i} both in the outskirts and spiral arms, why  is the gas at large galactocentric  distances not forming stars? 

\begin{figure}

\centering 
\includegraphics[ width=8 cm ]{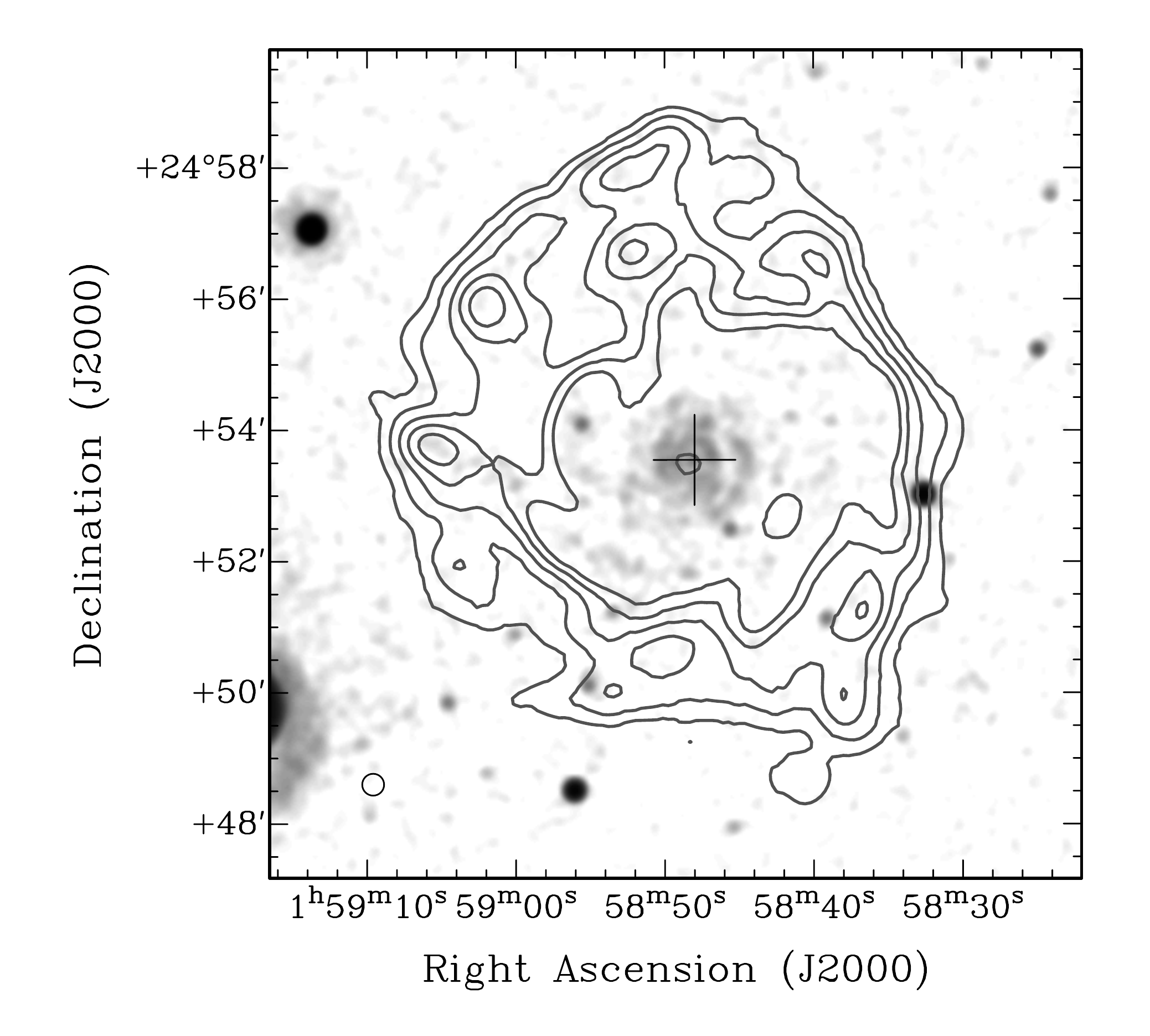}
\caption { \textit{GALEX} map smoothed to $10^{\prime\prime}$ resolution. In grey contours we represent the \textsc{H\,i} extent at a column density of 0.5, 1, 1.5, 2 and 2.5 $\times 10^{20}$ atoms cm$^{-2}$. The optical centre of the galaxy is marked by a cross. }
\label{galex}
\end{figure}

\textit{GALEX} observations (see Fig.~\ref{galex}) do not reveal any trace of a young stellar population in the outer  disc, implying that the conditions are not conducive to star formation. \citet{sch04} quotes a minimum gas column density necessary for a transition from warm atomic to cold molecular gas of around  3 to 10 \msun ~ pc$^{-2}$, equivalent to  4 -- 12 $ \times 10^{20}$ atoms cm$^{-2}$, assuming a model of a self--gravitating  disc containing dust and metals that is exposed to UV radiation. Similarly \citet{ler08} and \citet{big08} discuss the transition of  \textsc{H\,i}--to--H$_2$ on sub--kpc scales making use of the THINGS high spatial and velocity resolution data. They point to an \textsc{H\,i} surface brightness  limit of 9 \msun \,  pc$^{-2}$ or $\sim 10^{21}$ atoms cm$^{-2}$ found in all THINGS galaxies,  above which threshold molecular gas  dominates and hence star formation readily occurs.  

We find that column density peaks do not exceed  $\sim 3 \times 10^{20}$ atoms cm$^{-2}$ across the NGC\,765 \textsc{H\,i} outer disc, suggesting that little, if any, gas will have made the transition from neutral atomic to the molecular phase, a condition encountered more generally in spiral galaxies beyond R$_{25}$. NGC\,765 is simply a more extreme case in the sense that the outer \textsc{H\,i}  disc is much larger than in typical field spirals and resembles what has been found in other LSBs \citep{deblok98,auld06}. 
  
\section{Summary}
In this paper  we present multi--wavelength observations of NGC\,765. We highlight the more important results of this study:

\begin{itemize}
\item NGC\,765 presents a large \textsc{H\,i}  disc, 240\,kpc in diameter for an assumed distance of 72 Mpc. This is one of the largest \textsc{H\,i}  discs reported for a late type galaxy.

\item Assuming an optical scale length of 7.8\,kpc, NGC\,765 presents an \textsc{H\,i}--to--optical ratio of 15. This is considerably larger than what has been found in other giant LSBs such as Malin\,1.

\item The \textsc{H\,i}  disc morphology clearly presents two different regimes: an inner \textsc{H\,i} disk ($<50$\,kpc) tracing the spiral structure with the presence of at least three arms, and an outer, lower surface brightness plateau. This latter region is populated by several \textsc{H\,i} gas clumps of compact morphology.

\item The galaxy has two nearby companions: NGC\,765a, a previously  uncatalogued companion, towards the Northeast. Towards the South we have UGC\,1453, an irregular galaxy. We report the detection of a faint optical counterpart for NGC\,765a.

\item The outer  disc of NGC\,765 is slightly warped. It is populated with compact \textsc{H\,i} regions with sizes of $\sim 10$\,kpc and  \textsc{H\,i} masses of $\sim 10^8$\msun. These are possibly remnants of past mergers events.

\item We detect 1.4 GHz radio--continuum emission from the NGC\,765 optical core. This emission is weak, with a flux density of 2.16 mJy  and a luminosity of L$_{\rm 1.4\,GHz} = 1.3 \times 10^{21}$ W Hz$^{-1}$.  No radio continuum emission is found related to the stellar or \textsc{H\,i} disc.

\item \textit{Chandra} X--ray data  reveal emission coincident with the galaxy nucleus. We derive a luminosity of $\rm L_X$(0.2--10 keV) $\approx 1.7 \times 10^{40}$ erg s$^{-1}$. The radio and X--ray emission is compatible with the nucleus being powered by a weak AGN, as also suggested by its optical nuclear emission lines.

\item Star formation activity in NGC\,765 is low, with no star formation detected in the outer disk, suggesting the gas is well below the threshold for star formation.

\end{itemize}

Higher resolution \textsc{H\,i} observations will be needed to more fully understand the origin, structure, and expected evolution of this truly extraordinary \textsc{H\,i}  disc.

\vspace{1cm}

\textit{Acknowledgments }

We thank the anonymous referee for comments which helped us to improve the current paper.
AP and PB are funded by a stipend  from the University of Hertfordshire. AU has been supported via a Post Doctoral Research Assistantship and  ED is funded through a studentship, both from the UK  Science \& Technology Facilities Council.

The Digitized Sky Surveys were produced at the Space Telescope Science Institute under  government grant NAG W-2166. The images of these surveys are based on photographic data obtained on the Palomar Mountain and UK Schmidt Telescope. The plates were processed into  the present compressed digital form with the permission of these institutions. This research has made use of the NASA/IPAC Extragalactic Database (NED) which is operated by the Jet Propulsion Laboratory, California Institute of Technology, under contract with the National Aeronautics and Space Administration. We also acknowledge the usage of the HyperLeda database (http://leda.univ-lyon1.fr).

\bibliographystyle{mn2e}
\bibliography{giant}

\begin{thebibliography}{}

\bibitem[\protect\citeauthoryear{{Auld}, {de Blok}, {Bell} \& {Davies}}{{Auld}
  et~al.}{2006}]{auld06}
{Auld} R.,  {de Blok} W.~J.~G.,  {Bell} E.,    {Davies} J.~I.,  2006, \mnras,
  366, 1475

\bibitem[\protect\citeauthoryear{{Becker}, {White} \& {Helfand}}{{Becker}
  et~al.}{1995}]{bec95}
{Becker} R.~H.,  {White} R.~L.,    {Helfand} D.~J.,  1995, \apj, 450, 559

\bibitem[\protect\citeauthoryear{{Begeman}}{{Begeman}}{1987}]{beg87}
{Begeman} K.~G.,  1987, PhD thesis, Rijksuniversiteit Groningen, (1987)

\bibitem[\protect\citeauthoryear{{Bigiel}, {Leroy}, {Walter}, {Brinks}, {de
  Blok}, {Madore} \& {Thornley}}{{Bigiel} et~al.}{2008}]{big08}
{Bigiel} F.,  {Leroy} A.,  {Walter} F.,  {Brinks} E.,  {de Blok} W.~J.~G.,
  {Madore} B.,    {Thornley} M.~D.,  2008, \aj, 136, 2846

\bibitem[\protect\citeauthoryear{{Boomsma}, {Oosterloo}, {Fraternali}, {van der
  Hulst} \& {Sancisi}}{{Boomsma} et~al.}{2008}]{boo08}
{Boomsma} R.,  {Oosterloo} T.~A.,  {Fraternali} F.,  {van der Hulst} J.~M.,
  {Sancisi} R.,  2008, \aap, 490, 555

\bibitem[\protect\citeauthoryear{{Bothun}, {Impey} \& {McGaugh}}{{Bothun}
  et~al.}{1997}]{bot97}
{Bothun} G.,  {Impey} C.,    {McGaugh} S.,  1997, \pasp, 109, 745

\bibitem[\protect\citeauthoryear{{Bothun}, {Impey}, {Malin} \&
  {Mould}}{{Bothun} et~al.}{1987}]{bot87}
{Bothun} G.~D.,  {Impey} C.~D.,  {Malin} D.~F.,    {Mould} J.~R.,  1987, \aj,
  94, 23

\bibitem[\protect\citeauthoryear{{Bothun}, {Schombert}, {Impey} \&
  {Schneider}}{{Bothun} et~al.}{1990}]{bot90}
{Bothun} G.~D.,  {Schombert} J.~M.,  {Impey} C.~D.,    {Schneider} S.~E.,
  1990, \apj, 360, 427

\bibitem[\protect\citeauthoryear{{Bournaud}, {Combes}, {Jog} \&
  {Puerari}}{{Bournaud} et~al.}{2005}]{bou05}
{Bournaud} F.,  {Combes} F.,  {Jog} C.~J.,    {Puerari} I.,  2005, \aap, 438,
  507

\bibitem[\protect\citeauthoryear{{Braun}, {Thilker} \& {Walterbos}}{{Braun}
  et~al.}{2003}]{bra03}
{Braun} R.,  {Thilker} D.,    {Walterbos} R.~A.~M.,  2003, \aap, 406, 829

\bibitem[\protect\citeauthoryear{{Condon}, {Cotton}, {Greisen}, {Yin},
  {Perley}, {Taylor} \& {Broderick}}{{Condon} et~al.}{1998}]{con98}
{Condon} J.~J.,  {Cotton} W.~D.,  {Greisen} E.~W.,  {Yin} Q.~F.,  {Perley}
  R.~A.,  {Taylor} G.~B.,    {Broderick} J.~J.,  1998, \aj, 115, 1693

\bibitem[\protect\citeauthoryear{{Dalcanton}, {Spergel} \&
  {Summers}}{{Dalcanton} et~al.}{1997}]{dal97}
{Dalcanton} J.~J.,  {Spergel} D.~N.,    {Summers} F.~J.,  1997, \apj, 482, 659

\bibitem[\protect\citeauthoryear{{Das}, {Reynolds}, {Vogel}, {McGaugh} \&
  {Kantharia}}{{Das} et~al.}{2009}]{das09}
{Das} M.,  {Reynolds} C.~S.,  {Vogel} S.~N.,  {McGaugh} S.~S.,    {Kantharia}
  N.~G.,  2009, \apj, 693, 1300

\bibitem[\protect\citeauthoryear{{de Blok} \& {van der Hulst}}{{de Blok} \&
  {van der Hulst}}{1998}]{deblok98}
{de Blok} W.~J.~G.,  {van der Hulst} J.~M.,  1998, \aap, 336, 49

\bibitem[\protect\citeauthoryear{{de Jong}}{{de Jong}}{1996}]{dejong96}
{de Jong} R.~S.,  1996, \aaps, 118, 557

\bibitem[\protect\citeauthoryear{{de Jong} \& {van der Kruit}}{{de Jong} \&
  {van der Kruit}}{1994}]{dej94}
{de Jong} R.~S.,  {van der Kruit} P.~C.,  1994, \aaps, 106, 451

\bibitem[\protect\citeauthoryear{{de Vaucouleurs}, {de Vaucouleurs}, {Corwin}
  Jr., {Buta}, {Paturel} \& {Fouque}}{{de Vaucouleurs} et~al.}{1991}]{deVau91}
{de Vaucouleurs} G.,  {de Vaucouleurs} A.,  {Corwin} Jr. H.~G.,  {Buta} R.~J.,
  {Paturel} G.,    {Fouque} P.,  1991, {Third Reference Catalogue of Bright
  Galaxies}.
Springer, New York, NY (USA)

\bibitem[\protect\citeauthoryear{{Desroches} \& {Ho}}{{Desroches} \&
  {Ho}}{2009}]{des09}
{Desroches} L.-B.,  {Ho} L.~C.,  2009, \apj, 690, 267

\bibitem[\protect\citeauthoryear{{Disney} \& {Phillipps}}{{Disney} \&
  {Phillipps}}{1987}]{dis87}
{Disney} M.,  {Phillipps} S.,  1987, \nat, 329, 203

\bibitem[\protect\citeauthoryear{{Erwin}}{{Erwin}}{2005}]{erw05}
{Erwin} P.,  2005, \mnras, 364, 283

\bibitem[\protect\citeauthoryear{{Fall} \& {Efstathiou}}{{Fall} \&
  {Efstathiou}}{1980}]{fal80}
{Fall} S.~M.,  {Efstathiou} G.,  1980, \mnras, 193, 189

\bibitem[\protect\citeauthoryear{{Filho}, {Barthel} \& {Ho}}{{Filho}
  et~al.}{2006}]{fil06}
{Filho} M.~E.,  {Barthel} P.~D.,    {Ho} L.~C.,  2006, \aap, 451, 71

\bibitem[\protect\citeauthoryear{{Freeman} \& {Bland--Hawthorn}}{{Freeman} \&
  {Bland--Hawthorn}}{2002}]{fre02}
{Freeman} K.,  {Bland--Hawthorn} J.,  2002, \araa, 40, 487

\bibitem[\protect\citeauthoryear{{Freeman}}{{Freeman}}{1970}]{free70}
{Freeman} K.~C.,  1970, \apj, 160, 811

\bibitem[\protect\citeauthoryear{{Garcia}}{{Garcia}}{1993}]{gar93}
{Garcia} A.~M.,  1993, \aaps, 100, 47

\bibitem[\protect\citeauthoryear{{Giovanelli}, {Myers}, {Roth} \&
  {Haynes}}{{Giovanelli} et~al.}{1986}]{gio86}
{Giovanelli} R.,  {Myers} S.~T.,  {Roth} J.,    {Haynes} M.~P.,  1986, \aj, 92,
  250

\bibitem[\protect\citeauthoryear{{Governato}, {Brook}, {Mayer}, {Brooks},
  {Rhee}, {Wadsley}, {Jonsson}, {Willman}, {Stinson}, {Quinn} \&
  {Madau}}{{Governato} et~al.}{2010}]{gov10}
{Governato} F.,  {Brook} C.,  {Mayer} L.,  {Brooks} A.,  {Rhee} G.,  {Wadsley}
  J.,  {Jonsson} P.,  {Willman} B.,  {Stinson} G.,  {Quinn} T.,    {Madau} P.,
  2010, \nat, 463, 203

\bibitem[\protect\citeauthoryear{{Governato}, {Mayer}, {Wadsley}, {Gardner},
  {Willman}, {Hayashi}, {Quinn}, {Stadel} \& {Lake}}{{Governato}
  et~al.}{2004}]{gov04}
{Governato} F.,  {Mayer} L.,  {Wadsley} J.,  {Gardner} J.~P.,  {Willman} B.,
  {Hayashi} E.,  {Quinn} T.,  {Stadel} J.,    {Lake} G.,  2004, \apj, 607, 688

\bibitem[\protect\citeauthoryear{{Governato}, {Willman}, {Mayer}, {Brooks},
  {Stinson}, {Valenzuela}, {Wadsley} \& {Quinn}}{{Governato}
  et~al.}{2007}]{gov07}
{Governato} F.,  {Willman} B.,  {Mayer} L.,  {Brooks} A.,  {Stinson} G.,
  {Valenzuela} O.,  {Wadsley} J.,    {Quinn} T.,  2007, \mnras, 374, 1479

\bibitem[\protect\citeauthoryear{{Ho}, {Feigelson}, {Townsley}, {Sambruna},
  {Garmire}, {Brandt}, {Filippenko}, {Griffiths}, {Ptak} \& {Sargent}}{{Ho}
  et~al.}{2001}]{Ho01}
{Ho} L.~C.,  {Feigelson} E.~D.,  {Townsley} L.~K.,  {Sambruna} R.~M.,
  {Garmire} G.~P.,  {Brandt} W.~N.,  {Filippenko} A.~V.,  {Griffiths} R.~E.,
  {Ptak} A.~F.,    {Sargent} W.~L.~W.,  2001, \apjl, 549, L51

\bibitem[\protect\citeauthoryear{{Ho}, {Filippenko} \& {Sargent}}{{Ho}
  et~al.}{1997}]{ho97}
{Ho} L.~C.,  {Filippenko} A.~V.,    {Sargent} W.~L.~W.,  1997, \apjs, 112

\bibitem[\protect\citeauthoryear{{Hoffman}, {Silk} \& {Wyse}}{{Hoffman}
  et~al.}{1992}]{hof92}
{Hoffman} Y.,  {Silk} J.,    {Wyse} R.~F.~G.,  1992, \apjl, 388, L13

\bibitem[\protect\citeauthoryear{{Ibata}, {Gilmore} \& {Irwin}}{{Ibata}
  et~al.}{1994}]{iba94}
{Ibata} R.~A.,  {Gilmore} G.,    {Irwin} M.~J.,  1994, \nat, 370, 194

\bibitem[\protect\citeauthoryear{{Ibata} \& {Razoumov}}{{Ibata} \&
  {Razoumov}}{1998}]{iba98}
{Ibata} R.~A.,  {Razoumov} A.~O.,  1998, \aap, 336, 130

\bibitem[\protect\citeauthoryear{{Impey} \& {Bothun}}{{Impey} \&
  {Bothun}}{1989}]{imp89}
{Impey} C.,  {Bothun} G.,  1989, \apj, 341, 89

\bibitem[\protect\citeauthoryear{{Impey} \& {Bothun}}{{Impey} \&
  {Bothun}}{1997}]{imp97}
{Impey} C.,  {Bothun} G.,  1997, \araa, 35, 267

\bibitem[\protect\citeauthoryear{{Kalberla}, {Burton}, {Hartmann}, {Arnal},
  {Bajaja}, {Morras} \& {P{\"o}ppel}}{{Kalberla} et~al.}{2005}]{kal05}
{Kalberla} P.~M.~W.,  {Burton} W.~B.,  {Hartmann} D.,  {Arnal} E.~M.,  {Bajaja}
  E.,  {Morras} R.,    {P{\"o}ppel} W.~G.~L.,  2005, \aap, 440, 775

\bibitem[\protect\citeauthoryear{{Leroy}, {Walter}, {Brinks}, {Bigiel}, {de
  Blok}, {Madore} \& {Thornley}}{{Leroy} et~al.}{2008}]{ler08}
{Leroy} A.~K.,  {Walter} F.,  {Brinks} E.,  {Bigiel} F.,  {de Blok} W.~J.~G.,
  {Madore} B.,    {Thornley} M.~D.,  2008, \aj, 136, 2782

\bibitem[\protect\citeauthoryear{{Mapelli}, {Moore}, {Giordano}, {Mayer},
  {Colpi}, {Ripamonti} \& {Callegari}}{{Mapelli} et~al.}{2008}]{map08}
{Mapelli} M.,  {Moore} B.,  {Giordano} L.,  {Mayer} L.,  {Colpi} M.,
  {Ripamonti} E.,    {Callegari} S.,  2008, \mnras, 383, 230

\bibitem[\protect\citeauthoryear{{Mart{\'{\i}}nez\--Delgado}, {Aparicio},
  {G{\'o}mez-Flechoso} \& {Carrera}}{{Mart{\'{\i}}nez\--Delgado}
  et~al.}{2001}]{mar01}
{Mart{\'{\i}}nez\--Delgado} D.,  {Aparicio} A.,  {G{\'o}mez-Flechoso}
  M.~{\'A}.,    {Carrera} R.,  2001, \apjl, 549, L199

\bibitem[\protect\citeauthoryear{{McGaugh} \& {de Blok}}{{McGaugh} \& {de
  Blok}}{1997}]{mcg97}
{McGaugh} S.~S.,  {de Blok} W.~J.~G.,  1997, \apj, 481, 689

\bibitem[\protect\citeauthoryear{{Mihos}, {Spaans} \& {McGaugh}}{{Mihos}
  et~al.}{1999}]{mih99}
{Mihos} J.~C.,  {Spaans} M.,    {McGaugh} S.~S.,  1999, \apj, 515, 89

\bibitem[\protect\citeauthoryear{{Navarro}, {Frenk} \& {White}}{{Navarro}
  et~al.}{1995}]{nav95}
{Navarro} J.~F.,  {Frenk} C.~S.,    {White} S.~D.~M.,  1995, \mnras, 275, 56

\bibitem[\protect\citeauthoryear{{Noguchi}}{{Noguchi}}{2001}]{nog01}
{Noguchi} M.,  2001, \mnras, 328, 353

\bibitem[\protect\citeauthoryear{{O'Neil}, {Bothun}, {van Driel} \& {Monnier
  Ragaigne}}{{O'Neil} et~al.}{2004}]{one04}
{O'Neil} K.,  {Bothun} G.,  {van Driel} W.,    {Monnier Ragaigne} D.,  2004,
  \aap, 428, 823

\bibitem[\protect\citeauthoryear{{Pe{\~n}arrubia}, {McConnachie} \&
  {Babul}}{{Pe{\~n}arrubia} et~al.}{2006}]{pen06}
{Pe{\~n}arrubia} J.,  {McConnachie} A.,    {Babul} A.,  2006, \apjl, 650, L33

\bibitem[\protect\citeauthoryear{{Pickering}, {Impey}, {van Gorkom} \&
  {Bothun}}{{Pickering} et~al.}{1997}]{pic97}
{Pickering} T.~E.,  {Impey} C.~D.,  {van Gorkom} J.~H.,    {Bothun} G.~D.,
  1997, \aj, 114, 1858

\bibitem[\protect\citeauthoryear{{Pisano} \& {Wilcots}}{{Pisano} \&
  {Wilcots}}{2003}]{pis03}
{Pisano} D.~J.,  {Wilcots} E.~M.,  2003, \apj, 584, 228

\bibitem[\protect\citeauthoryear{{Rahman}, {Howell}, {Helou}, {Mazzarella} \&
  {Buckalew}}{{Rahman} et~al.}{2007}]{rah07}
{Rahman} N.,  {Howell} J.~H.,  {Helou} G.,  {Mazzarella} J.~M.,    {Buckalew}
  B.,  2007, \apj, 663, 908

\bibitem[\protect\citeauthoryear{{Roberts} \& {Haynes}}{{Roberts} \&
  {Haynes}}{1994}]{rob94}
{Roberts} M.~S.,  {Haynes} M.~P.,  1994, \araa, 32, 115

\bibitem[\protect\citeauthoryear{{Sancisi} \& {Fraternali}}{{Sancisi} \&
  {Fraternali}}{2007}]{san07}
{Sancisi} R.,  {Fraternali} F.,  2007, ArXiv e-prints

\bibitem[\protect\citeauthoryear{{Sancisi}, {Fraternali}, {Oosterloo} \& {van
  der Hulst}}{{Sancisi} et~al.}{2008}]{san08}
{Sancisi} R.,  {Fraternali} F.,  {Oosterloo} T.,    {van der Hulst} T.,  2008,
  \aapr, 15, 189

\bibitem[\protect\citeauthoryear{{Scannapieco}, {Tissera}, {White} \&
  {Springel}}{{Scannapieco} et~al.}{2008}]{scann08}
{Scannapieco} C.,  {Tissera} P.~B.,  {White} S.~D.~M.,    {Springel} V.,  2008,
  \mnras, 389, 1137

\bibitem[\protect\citeauthoryear{{Schaye}}{{Schaye}}{2004}]{sch04}
{Schaye} J.,  2004, \apj, 609, 667

\bibitem[\protect\citeauthoryear{{Schneider}, {Thuan}, {Magri} \&
  {Wadiak}}{{Schneider} et~al.}{1990}]{sch90}
{Schneider} S.~E.,  {Thuan} T.~X.,  {Magri} C.,    {Wadiak} J.~E.,  1990,
  \apjs, 72, 245

\bibitem[\protect\citeauthoryear{{Schombert}}{{Schombert}}{1998}]{schombert98}
{Schombert} J.,  1998, \aj, 116, 1650

\bibitem[\protect\citeauthoryear{{Schulman}, {Bregman} \& {Roberts}}{{Schulman}
  et~al.}{1994}]{shu94}
{Schulman} E.,  {Bregman} J.~N.,    {Roberts} M.~S.,  1994, \apj, 423, 180

\bibitem[\protect\citeauthoryear{{Sommer-Larsen}, {Gelato} \&
  {Vedel}}{{Sommer-Larsen} et~al.}{1999}]{som99}
{Sommer-Larsen} J.,  {Gelato} S.,    {Vedel} H.,  1999, \apj, 519, 501

\bibitem[\protect\citeauthoryear{{Sommer-Larsen}, {G{\"o}tz} \&
  {Portinari}}{{Sommer-Larsen} et~al.}{2003}]{som03}
{Sommer-Larsen} J.,  {G{\"o}tz} M.,    {Portinari} L.,  2003, \apj, 596, 47

\bibitem[\protect\citeauthoryear{{Sprayberry}, {Impey}, {Bothun} \&
  {Irwin}}{{Sprayberry} et~al.}{1995}]{spr95}
{Sprayberry} D.,  {Impey} C.~D.,  {Bothun} G.~D.,    {Irwin} M.~J.,  1995, \aj,
  109, 558

\bibitem[\protect\citeauthoryear{{Sprayberry}, {Impey}, {Irwin}, {McMahon} \&
  {Bothun}}{{Sprayberry} et~al.}{1993}]{spr93}
{Sprayberry} D.,  {Impey} C.~D.,  {Irwin} M.~J.,  {McMahon} R.~G.,    {Bothun}
  G.~D.,  1993, \apj, 417, 114

\bibitem[\protect\citeauthoryear{{Springel}, {White}, {Jenkins}, {Frenk},
  {Yoshida}, {Gao}, {Navarro}, {Thacker}, {Croton}, {Helly}, {Peacock}, {Cole},
  {Thomas}, {Couchman}, {Evrard}, {Colberg} \& {Pearce}}{{Springel}
  et~al.}{2005}]{spr05}
{Springel} V.,  {White} S.~D.~M.,  {Jenkins} A.,  {Frenk} C.~S.,  {Yoshida} N.,
   {Gao} L.,  {Navarro} J.,  {Thacker} R.,  {Croton} D.,  {Helly} J.,
  {Peacock} J.~A.,  {Cole} S.,  {Thomas} P.,  {Couchman} H.,  {Evrard} A.,
  {Colberg} J.,    {Pearce} F.,  2005, \nat, 435, 629

\bibitem[\protect\citeauthoryear{{Springob}, {Haynes}, {Giovanelli} \&
  {Kent}}{{Springob} et~al.}{2005}]{springob05}
{Springob} C.~M.,  {Haynes} M.~P.,  {Giovanelli} R.,    {Kent} B.~R.,  2005,
  \apjs, 160, 149

\bibitem[\protect\citeauthoryear{{van der Hulst} \& {Sancisi}}{{van der Hulst}
  \& {Sancisi}}{2005}]{van05}
{van der Hulst} J.~M.,  {Sancisi} R.,  2005, in {R.~Braun} ed., Extra-Planar
  Gas Vol.~331 of Astronomical Society of the Pacific Conference Series, {Gas
  Accretion in Galactic Disks}.
p.~139

\bibitem[\protect\citeauthoryear{{van der Hulst}, {Terlouw}, {Begeman},
  {Zwitser} \& {Roelfsema}}{{van der Hulst} et~al.}{1992}]{vanhul92}
{van der Hulst} J.~M.,  {Terlouw} J.~P.,  {Begeman} K.~G.,  {Zwitser} W.,
  {Roelfsema} P.~R.,  1992, in {D.~M.~Worrall, C.~Biemesderfer, \& J.~Barnes}
  ed., Astronomical Data Analysis Software and Systems I Vol.~25 of
  Astronomical Society of the Pacific Conference Series, {The Groningen Image
  Processing SYstem, GIPSY}.
p.~131

\bibitem[\protect\citeauthoryear{{van Moorsel}}{{van Moorsel}}{1982}]{moor82}
{van Moorsel} G.~A.,  1982, \aap, 107, 66

\bibitem[\protect\citeauthoryear{{Walter}, {Brinks}, {de Blok}, {Bigiel},
  {Kennicutt}, {Thornley} \& {Leroy}}{{Walter} et~al.}{2008}]{wal08}
{Walter} F.,  {Brinks} E.,  {de Blok} W.~J.~G.,  {Bigiel} F.,  {Kennicutt}
  R.~C.,  {Thornley} M.~D.,    {Leroy} A.,  2008, \aj, 136, 2563

\bibitem[\protect\citeauthoryear{{Wilcots}}{{Wilcots}}{2001}]{wil01}
{Wilcots} E.~M.,  2001, in {J.~E.~Hibbard, M.~Rupen, \& J.~H.~van Gorkom} ed.,
  Gas and Galaxy Evolution Vol.~240 of Astronomical Society of the Pacific
  Conference Series, {The Extended HI Environment of Galaxies}.
p.~167

\end{thebibliography}

\appendix
\section{NGC\,765 channel maps}
\label{appendix-a}

\begin{figure*}
\begin{center}
\centering 
\includegraphics[trim=200 50 200 50, width=10 cm]{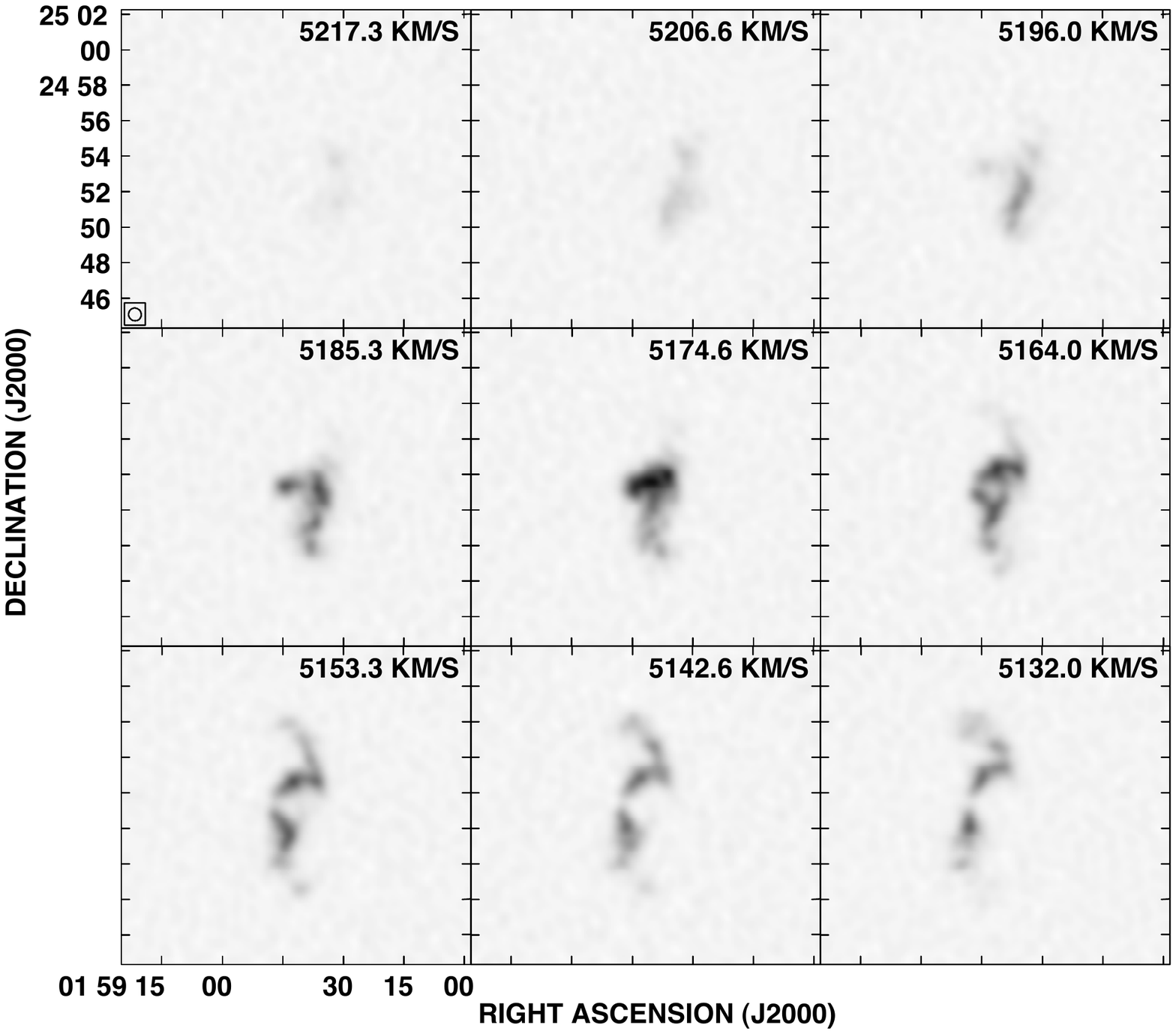}
\caption{Individual velocity channels of NGC\,765. The beam is shown in the left bottom corner of the top left panel. Heliocentric velocities are indicated in each panel.}
\label{765_CHAN_1}
\end{center}
\end{figure*}

In Fig.~\ref{765_CHAN_1} we present the \textsc{H\,i} channel maps of NGC\,765 as a set of grey--scale maps. They were generated using  {\sc imagr} choosing a \textsc{robust=0} weight in order to obtain the best compromise between resolution and sensitivity. The cube was cleaned to a flux threshold of $2\sigma$.

\setcounter{figure}{0}
\begin{figure*}
\begin{center}
\centering \includegraphics[trim=200 50 200 50,width=10 cm]{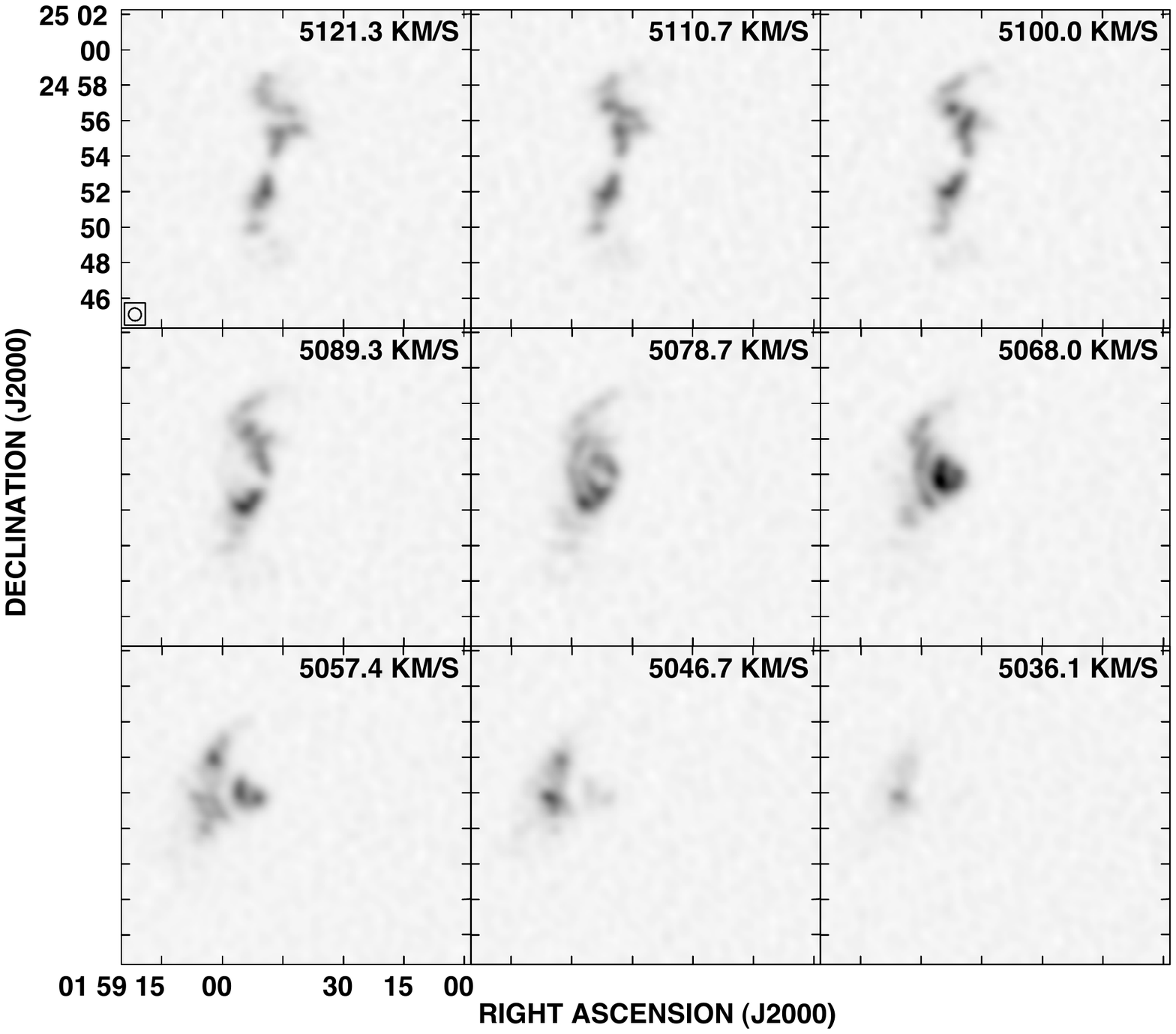}
\caption{Individual velocity channels of NGC\,765 (Continued). }
\label{765_CHAN_2}
\end{center}
\end{figure*}

\section{Radio--continuum sources}
\label{appendix-b}

\begin{figure*}
\centering 
\includegraphics[width=10 cm]{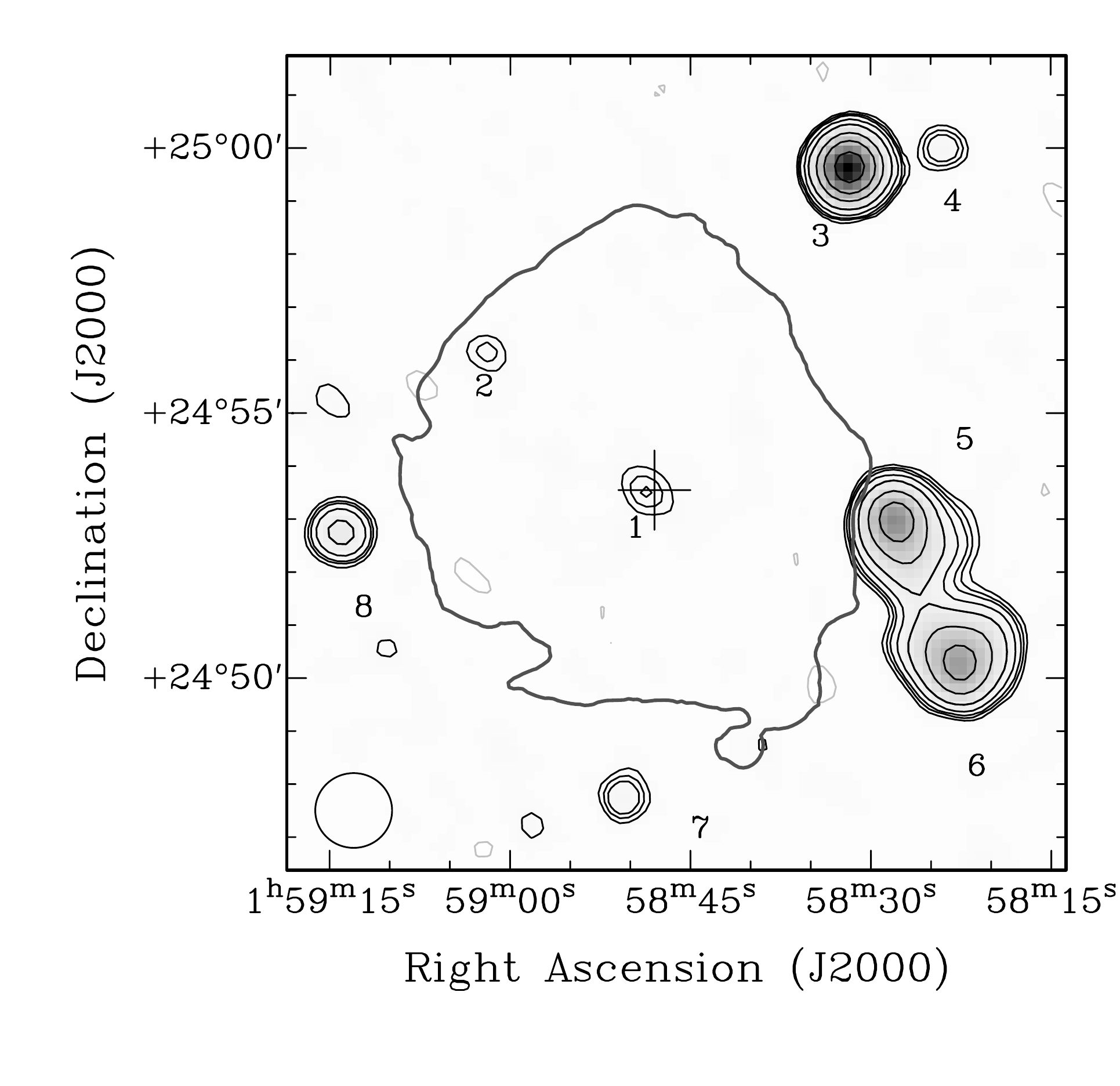}
\caption{ The radio continuum emission of the NGC\,765 field at 1.4 GHz. The contours are at (-2, 2, 3, 4.5, 6, 10, 20, 50, 100 and 200) times the rms of 0.15 mJy beam$^{-1}$. The negative contour is plotted as light grey. The extent of the \textsc{H\,i}  disc of  NGC\, 765 at a level of 5 $\times 10^{19}$ atoms cm$^{-2}$ is plotted with a single dark grey contour. The optical centre of the galaxy is represented by the cross. Source 1 coincides with the optical centre of NGC\,765.}
\label{continuum}
\end{figure*}

There are several radio--continuum sources in the field, in addition to the source coinciding with the optical nucleus. Except for continuum source number 2, they fall outside of the outline defined by the \textsc{H\,i}, as can be seen in Fig. \ref{continuum}. We present in Table~\ref{RADIO_CONTINUUM_TABLE} the properties of the radio--continuum sources in the field of NGC\,765. Of the eight sources detected above 4.5$\sigma$,  sources 1 and 2 are  found to fall within the \textsc{H\,i}  disc. Source 2 coincides to within $17^{\prime\prime}$, or 5\,kpc, with \textsc{H\,i} compact region B (see Sect. \ref{HIcontent}). Sources 4, 6, 7  and 8 were all detected by the NVSS as well. Source 3 has two 2MASS objects identified as possible counterpart. Source 4 was identified by 2MASS as a galaxy. The remaining sources have no NVSS detection or NED identification.

\begin{table*}
\caption{List of radio continuum sources detected in the vicinity of NGC\,765.}
\centering
\begin{tabular}{ccccccccccc}
\hline
\hline 
Source & $\alpha$ & $\delta$ & Deconvolved size & F$\rm _P$ & F$\rm _I$  & PA & S/N & Type & Identification  \\
& (J2000) & (J2000) & \arcsec$\times$\arcsec & mJy  &mJy &  \degr &  &  &\\
\hline
1 & $01^h\, 58^m\, 48.5^s$    & 24\degr~53\arcmin~39\arcsec  & $61 \times 34$ & 0.83  & 2.16 & 58 & 14 & G  & NGC\,765\\
2 & $01^h\, 59^m\, 01.9^s$    & 24\degr~56\arcmin~06\arcsec  & $23 \times 14 $ &  0.78   & 0.84  & 46 & 6 & \ldots & \ldots \\
3 & $01^h\, 58^m\, 31.7^s$    & 24\degr~59\arcmin~39\arcsec  & $17\times 9 $ &  43.3   & 47.1  & 157 & 314 & G, RadioS & 2MASX J01583181+2459391 \\
 & & & & & & & & & 2MASX J01582855+2459367 \\
 & & & & & & & & & NVSS J015831+245938 \\ 
4 & $01^h\, 58^m\, 24.1^s$    & 24\degr~59\arcmin~59\arcsec  & $6\times 0$ &  1.38  & 1.18  & 111 & 8 & G  & 2MASX J01582399+2459598 \\
5 & $01^h\, 58^m\, 27.7^s$    & 24\degr~52\arcmin~53\arcsec  & $51\times 29 $ &  21.0  & 37.0  & 21 & 247 & RadioS & NVSS J015827+245252 \\
6 & $01^h\, 58^m\, 22.7^s$    & 24\degr~50\arcmin~22\arcsec  & $52\times 43$ &  18.8  & 40.3  & 37 & 269 & RadioS & NVSS J015822+245022 \\
7 & $01^h\, 58^m\, 50.5^s$    & 24\degr~47\arcmin~46\arcsec  & $17\times 13$ &  1.44  & 1.37  & 36 & 9 & \ldots &\ldots \\
8 & $01^h\, 59^m\, 14.1^s$    & 24\degr~52\arcmin~44\arcsec  & $12\times 7$ &  4.03  & 3.97  & 137 & 27 & RadioS & NVSS J015914+245248 \\
\hline
\hline
\\
\end{tabular}
\label{RADIO_CONTINUUM_TABLE}
\hspace{1cm}
\begin{flushleft}
Notes: Column 1: Source number; Column 2 \& 3: Right ascension and declination; Column 4: Deconvolved source size; Column 5: Peak flux; Column 6: Integrated flux; Column 7: Position angle; Column 8: Signal--to--noise ratio; Column 9: object type --- G denotes galaxy and RadioS radio source; Column 10: Identification.   
\end{flushleft}     
\end{table*}

\section{X--ray sources}
\label{appendix-c}

The data reduction of the Chandra archival observations of NGC\,765 revealed a total of six sources in the general field, one coinciding with the nucleus. Because the photon statistics are low (see Table 4), we did not attempt to extract spectra, but instead opted to use the PIMMs tool (v3.9i) to convert source counts into fluxes. We assumed the appropriate Cycle 8 calibration curves for the ACIS-S detector, an input energy of 0.2-10 keV, a simple power law model with a weighted galactic absorption towards NGC\,765 of 8.03 $\times$ 10$^{20}$ cm$^{-2}$ \citep{kal05}, fixed photon index of 2.0 (appropriate for LLAGN and for X-ray binaries) and a redshift of z=0.017 (NED). The results are presented in Table~\ref{XRAY_TABLE}. 

\begin{figure*}
\centering 
\includegraphics[width=10 cm,angle=-90]{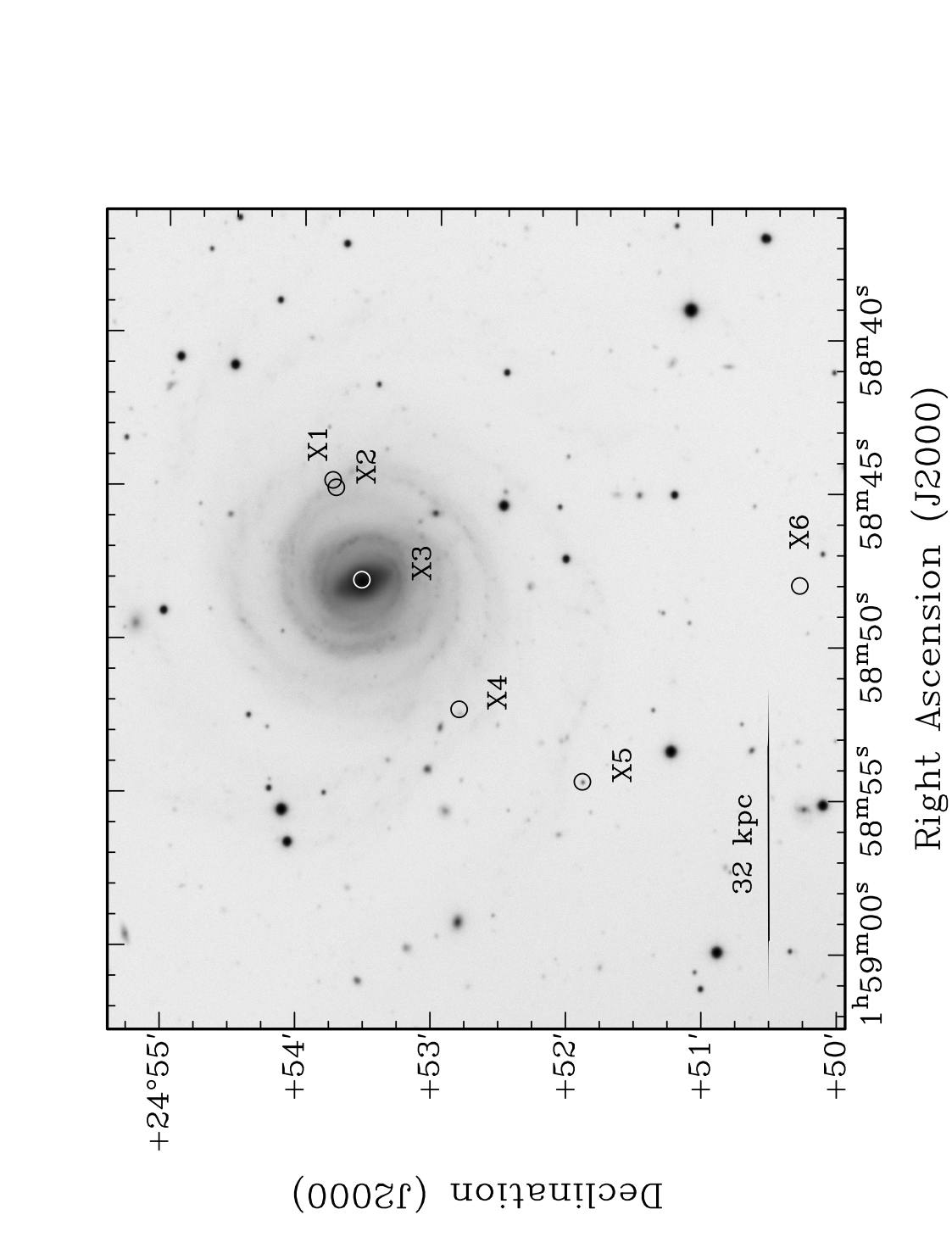}
\caption{The location of the detected Chandra X--ray sources overlaid on the optical INT map of NGC\,765. The central source X3 coincides  with the optical centre and the central radio--continuum source of  NGC\,765. }
\label{xray_figure}
\end{figure*}

X--ray source X3 (Table \ref{XRAY_TABLE} and Fig.~\ref{xray_figure}) is  relatively  compact, coincides with the optical nucleus of NGC\,765 and with the  nuclear radio--continuum source (Table \ref{RADIO_CONTINUUM_TABLE}). It has an X--ray luminosity in the  0.2--10 keV energy band of L$_{\rm X} \approx 1.7 \times 10^{40}$ erg s$^{-1}$. This low X--ray luminosity is typical of low--luminosity AGN \citep{Ho01}. 

Due to the low X--ray counts and lack of temporal information, the nature of the remaining X--ray sources in the field can not be established with certainty. Sources X4 and X5 (Fig.~\ref{xray_figure}) coincide spatially with stellar counterparts but are unidentified in both the NED and 2MASS catalogues. These could be background quasars but because of a lack of redshift information, we  estimated their X--ray luminosities assuming they are at the distance of NGC\,765, although there is no \textit{a priori} evidence that they belong to this system. The low--luminosity and the off--nuclear position of sources X1, X2 and X6 suggest these are low luminosity X--ray binaries. Their luminosities have been estimated again assuming they are at  the redshift of NGC\,765. 

In order to detect any hypothetical extended, diffuse X--ray emission, we subtracted the point sources from the image and interpolated the gaps with appropriate values for the background, using the statistics of the surrounding pixels. We then smoothed the image to a 5 arcsec beam. Contrary to \citep{das09} who find a diffuse gas component associated with the galaxy centre and confined to the bulge at a luminosity of 5.5 $\times$ 10$^{39}$ erg s$^{-1}$, we find no significant diffuse X--ray emission anywhere within the \textsc{H\,i} disc.

\begin{table*}
\caption{Chandra X--ray detections in the NGC\,765 field.}
\centering
\begin{tabular}{ccccccccc}
\hline
\hline 
source & $\alpha$  &  $\delta$ &  N$\rm _S$   &    N$\rm _{bg}$  & S/N & Unabsorbed flux & Luminosity \\
& J(2000) & J(2000) &  &  & & 10$^{-14}$ erg s$^{-1}$ cm$^{-2}$ & 10$^{40}$ erg s$^{-1}$ \\
\hline
&&&&&& 1.6 & 1.0  \\
X1   &$1^h\, 58^m\, 44.7^s$ & 24\degr\,53\arcmin\, 46.1\arcsec &    6.4   &      1.5 & 4.5   & 0.9 & 0.6  \\
&&&&&& 0.6 & 0.4 \\
\\
&&&&&& 3.7 & 2.3 \\
X2  & $1^h\, 58^m\, 44.9^s$  & 24\degr\,53\arcmin\,44.6\arcsec &   15.2   &     2.6  & 5.8  & 2.2 & 1.3 \\
&&&&&&  1.5 & 0.9\\
\\
&&&&&& 2.8 & 1.7\\
X3   & $1^h\, 58^m\, 48.0^s$ & 24\degr\,53\arcmin\,33.0\arcsec &  11.8   &  3.4   &3.5  & 1.7 & 1.0   \\
&&&&&& 1.2 & 0.7  \\
\\
&&&&&& 1.8 & 1.1 \\
X4   & $1^h\, 58^m\, 52.3^s$  & 24\degr\,52\arcmin\,48.5\arcsec & 7.5     &      2.1  & 3.6 & 1.1 & 0.7 \\
&&&&&& 0.8 & 0.5 \\
\\
&&&&&& 2.8 & 1.7\\
X5 & $1^h\, 58^m\, 54.5^s$ & 24\degr\,51\arcmin\,53.9\arcsec   &   11.5   &      3.7      & 3.1&  1.6 & 1.0 \\
&&&&&& 1.1 & 0.7 \\
\\
&&&&&& 3.3 & 2.0 \\
X6 & $1^h\, 58^m\, 48.1^s$ & 24\degr\,50\arcmin\,19.4\arcsec  &    13.5    &     4.8   & 3.0  & 1.9 & 1.2 \\
&&&&&& 1.3 & 0.8  \\
\hline
\hline
\end{tabular}
\label{XRAY_TABLE}
\hspace{1cm}
\begin{flushleft}
Notes: Column 1: Source name; Column 2 \& 3: Right ascension and declination; Column 4: Source counts; Column 5: Background counts; Column 6: Signal--to--noise ratio; Column 7: Unabsorbed X--ray flux in the energy bands: 0.2 -- 10 keV (total; top row), 0.2 -- 2 keV (soft;  middle row) and  2 -- 10 keV (hard; bottom row); Column 8: X--ray luminosity in the same energy bands, assuming a distance of 72\,Mpc.
\end{flushleft}   

\end{table*}

\label{lastpage}

\end{document}